\documentclass[12pt]{article}

\usepackage{amsmath}
\usepackage{amssymb}

\usepackage{hyperref}

\usepackage{graphicx}
\graphicspath{{./img/}{./}}
\usepackage{caption,subcaption}

\usepackage{booktabs}

\newenvironment{sciabstract}{%
\begin{quote} \bf}
{\end{quote}}

\usepackage{color}
\usepackage[normalem]{ulem}

\newcommand{\del}[1]{{\sout{#1}}}

\usepackage{cancel}

\renewcommand{\del}[1]{{}}   

\title{Doping the holographic Mott insulator.}

\author
{Tomas Andrade$^{2,3}$, Alexander Krikun$^{1\ast}$, Koenraad Schalm$^{1}$ \& Jan Zaanen$^{1}$ \\
 \normalsize{$^{1}$Institute-Lorentz for Theoretical Physics,~
Leiden University}\\
\normalsize{P.O. Box 9506, 2300 RA Leiden, The Netherlands}\\
\\
\normalsize{$^{2}$Rudolf Peierls Centre for Theoretical Physics,~
University of Oxford}\\
\normalsize{1 Keble Road, Oxford OX1 3NP, UK}\\
\\
\normalsize{$^{3}$
Departament de F\'isica Qu\`antica i Astrof\'isica,}\\
\normalsize{Institut de Ci\`encies del Cosmos (ICCUB), Universitat de Barcelona} \\
\normalsize{Mart\'i i Franqu\`es 1, E-08028 Barcelona, Spain}\\ 
\\
\normalsize{$^\ast$To whom correspondence should be addressed; E-mail:  krikun@lorentz.leidenuniv.nl.}
}

\begin{document} 

\date{}
\maketitle





 \baselineskip18pt



\begin{sciabstract}
Mott insulators form because of strong electron repulsions, being at the heart of strongly correlated electron physics. Conventionally 
these are understood as classical ``traffic jams'' of electrons described by a short-ranged entangled product ground state. 
Exploiting the holographic duality, which maps the physics of densely entangled matter onto gravitational black hole physics, we 
show how Mott-insulators can be constructed departing from entangled non-Fermi liquid metallic states, such as the strange metals found in cuprate superconductors. 
These ``entangled Mott insulators'' have traits in common with the ``classical'' Mott insulators, such as the formation of Mott gap in the optical conductivity,  super-exchange-like
interactions, and form ``stripes'' when doped.  They also exhibit new properties: 
the ordering wave vectors are detached from the number of electrons in  the unit cell, and the DC resistivity diverges algebraically instead of exponentially 
as function of temperature. 
These results may shed light on the mysterious ordering phenomena
observed in underdoped cuprates.
\end{sciabstract}

The ``hard'' Mott insulators (MI) realised in stoichiometric transition metals salts are regarded as one of the few entities that are well understood in the arena of strongly correlated electron systems \cite{zaanen1985band}.
The principles are given away by Hubbard-type models: given an integer number of electrons per unit
cell, any charge fluctuation gives rise to an excess local Coulomb energy ``$U$'' and when this becomes much larger than the bandwidth, quite literally a traffic jam of electrons is formed. 
This state can be adiabatically continued to the weak interaction limit using conventional (Hartree-Fock) mean-field theory, where it turns into a ``BCS-like'' commensurate spin density wave \cite{fradkin2013field}.
The language of quantum information reveals the key element: Hartree-Fock rests on the assumption that the ground state is a ``classical'' short ranged entangled product state \cite{wen2004quantum}.
At strong coupling the quantized single electron charges are just localized inside the unit cell. At weak coupling one has to accommodate Fermi-statistics, but the case can be made
precise that even the Fermi gas is a product state in momentum space \cite{zaanen2008pacifying}. The perfectly nested density wave (weak coupling MI)  then  ``inherits'' its lack of macroscopic entanglement
from the underlying Fermi-liquid.  
 
However, matter may also be ``truly quantum'' in the sense of quantum information: the vacuum state may be an irreducible coherent superposition involving an extensive part of the exponentially large many body Hilbert space. Little is known with certainty  given the quantum complexity: a quantum computer is needed to address it thwith confidence. Indications 
are accumulating that the strange metals realised in the cuprate high Tc superconductors may be of this kind \cite{Zaanen:2015oix}. Upon lowering temperature in the under-doped regime, this strange metal
becomes unstable towards a myriad of ``intertwined'' ordering phenomena that do depend critically on the ionic lattice potential \cite{keimer2015quantum,fradkin2015colloquium}. It has become increasingly clear that this pseudo-gap order 
does not seem explainable in terms of conventional mean-field language \cite{keimer2015quantum,mesaros2016commensurate}. Could it be that these ordering phenomena inherit the many-body entanglement of the strange metal? If so, do these submit 
to general emergence principles of a new kind that can be identified in experiment?  
 
A new mathematical machinery has become available which can address
this question at least to a degree.  There is strong evidence that the holographic duality \cite{ammon2015gauge}
(or AdS/CFT correspondence) discovered in string theory describes
generic properties of certain classes of such densely entangled
quantum matter \cite{Zaanen:2015oix}.  In particular,
holographic strange metals are emergent quantum critical
phases  that behave in key regards suggestively similar to the
laboratory strange metals  (local quantum criticality
\cite{faulkner2010strange,iqbal2011lectures}, Planckian dissipation
\cite{Policastro:2001yc,hartnoll2016holographic}).  Here we will
explore what holography has to say about the emergence of
``entangled Mott insulators''.  

The results reveal generalities which are intriguing and suggestive towards experiment.  On the one hand, the holographic realisation of the Mott insulator shows properties similar 
to the conventional variety. The optical conductivity has the
  similar characteristics as the inter Hubbard band transitions found in hard
Mott insulators  (Fig.\,1e) \cite{zaanen1990systematics,rozenberg1995optical} and an analogue of superexchange interaction \cite{anderson1950antiferromagnetism,zaanen1987electronic} can be identified (Fig.\,1f). Upon doping, close analogies of the ``spin stripes'' \cite{zaanen1989charged,tranquada1995evidence,vojta2009lattice,zheng2016stripe,huang2016numerical} are formed 
(Fig.\,4b).  However, they also reveal 
unconventional features reflecting their entangled nature. Reminescent of the charge order in cuprates, the periodicity of the charge order that forms upon doping displays commensurate plateaux (Fig.\,3b), staying constant in a range of doping levels. Most intriguingly, holography insists that charge cannot be truly localized as in a ``product state'' Mott insulator. Instead, a reconfigured  emergent quantum critical phase 
emerges at low energies characterized by a DC resistivity that increases algebraically instead of exponentially for decreasing temperature (Fig.\,5a). This may shine light on the long standing puzzle of the slow  (logarithmic)  rise of the resistivity in striped cuprates. 

Let us now discuss how we arrive at these results. Conventional Mott insulators obey the rule that the unit cell should contain an {\em integer} number of electrons in order for them to form. 
However, this implicitly rests on product state structure, and the information regarding the graininess of the microscopic electron charge is generically washed out in strongly entangled states including 
the ones described by holography. There is, however, a truly general
definition of a Mott insulator that circumvents the confines of
microscopic product states: {\em a Mott insulator is an electron crystal that is commensurately 
pinned by a periodic background potential.} A crystal formed in the Galilean continuum is a perfect metal since it can freely slide -- its massless longitudinal phonon is dual to a current that is protected by total 
momentum conservation. This sliding mode will acquire a pinning energy in a commensurate background lattice and this is the general meaning of a Mott gap. 

This is not a practical way to construct a Mott insulator departing from the electron gas at metallic densities in the Galilean continuum since this lacks a natural tendency to crystallize. 
Holographic strange metals on the other hand are known to have crystallisation tendencies, where the most natural form\cite{Donos:2013gda} intriguingly involves a most literal form of ``intertwined'' order similar to that observed in underdoped cuprates \cite{fradkin2015colloquium,keimer2015quantum,fauque2006magnetic,li2008unusual,li2010hidden,Zhao2017global,li2007two,rajasekaran2017probing,hamidian2016detection}. 
The AdS/CFT correspondence shows that the properties of quantum matter
can be computed in terms of a holographic gravitational ``dual'' in a space with one 
extra dimension \cite{ammon2015gauge,Zaanen:2015oix}. Strange metallic
states appear to be in one-to-one correspondence to charged black holes
in this gravitational system. It was discovered that topological
terms in the gravity theory (theta- and  Chern-Simons terms in even- and odd- dimensions) have the effect that the horizon of the black hole 
 becomes unstable towards spatial modulations
 \cite{Ooguri:2010kt,Donos:2011bh} at the ``expense'' that the charge
 modulation is ``intertwined'' \cite{Cai:2017qdz} with parity breaking and the emergence of
 spontaneous diamagnetic ``particle-hole'' currents.

Here we study this holographic crystallisation in the presence of an
external periodic potential. This demands advanced numerics to solve the gravitational problem;
our ``corrugated black holes'' are among the most involved solutions in stationary general relativity (GR). To keep the computations manageable we focus on simple harmonic background potentials and especially a unidirectional 
translational symmetry breaking.
We consider here specifically the minimal version of such a
gravitational theory \cite{Donos:2011bh}. The basis is Einstein-Maxwell
theory in 3+1 dimensions with a negative cosmological constant, 
 describing the simplest holographic strange metal in 2+1 dimensions.
 The crucial extra ingredient is the topological parity-odd $\vartheta$-term
 coupling the Maxwell field $A_{\mu}$ with field strength $F_{\mu\nu}$
 to a dynamical pseudoscalar field $\psi$, such that the action becomes,
\begin{equation}\label{S_0}
	S = \int d^4 x \sqrt{- g} \left( R - 2 \Lambda- \frac{1}{2} (\partial \psi)^2 - \frac{\tau(\psi)}{4} F^2 -  W(\psi) \right)
	 - \frac{1}{2} \int {\vartheta}(\psi) F \wedge F
\end{equation}
The qualitative features we reveal depend only mildly on the precise
form of the functions $\tau(\psi), W(\psi), \theta(\psi)$ (see
Supplementary Material).
The solution to the equations of motion will asymptote to
  anti-de-Sitter space (AdS), on the (conformal)
  boundary of which the dual theory lives. At finite density and temperature of the dual theory, 
  a charged (Reissner-Nordstrom) black hole 
is present in the deep interior, which  famously
translates to a locally quantum critical strange metal
(see Supplementary Material for the dictionary entries).  As
temperature is
lowered, the $\vartheta$-term 
causes the horizon to become unstable towards a modulation of the spatial geometry that breaks
translational symmetry spontaneously.  Here we choose the simplest version, corresponding 
to a unidirectional symmetry breaking in the ``$x$'' direction (see
\cite{Withers:2014sja, Cai:2017qdz} for a ``full'' 2D crystallisation).  This is driven by the condensation of the pseudoscalar $\psi$ 
representing a spontaneous breaking of parity on the boundary. The structure
of the  $\vartheta$-term makes that this is accompanied by
condensation of Maxwell field strength. This translates to the
formation of spontaneous currents running in the $y$ direction, while
a concomitant charge  density wave {(CDW)} develops. 
A reliable, consistent result is only obtained if one solves the full equations of motion in the gravitational theory, 
and this includes the change in geometry due to back-reaction. Given the inhomogeneous nature of the bulk space time,
this involves a considerable numerical GR effort since the Einstein equations represent a system of non-linear partial 
differential equations (see Supplementary Material). The result is represented in Fig.\,1b.

\begin{figure}[ht!]
\centering
 \includegraphics[width=1\linewidth]{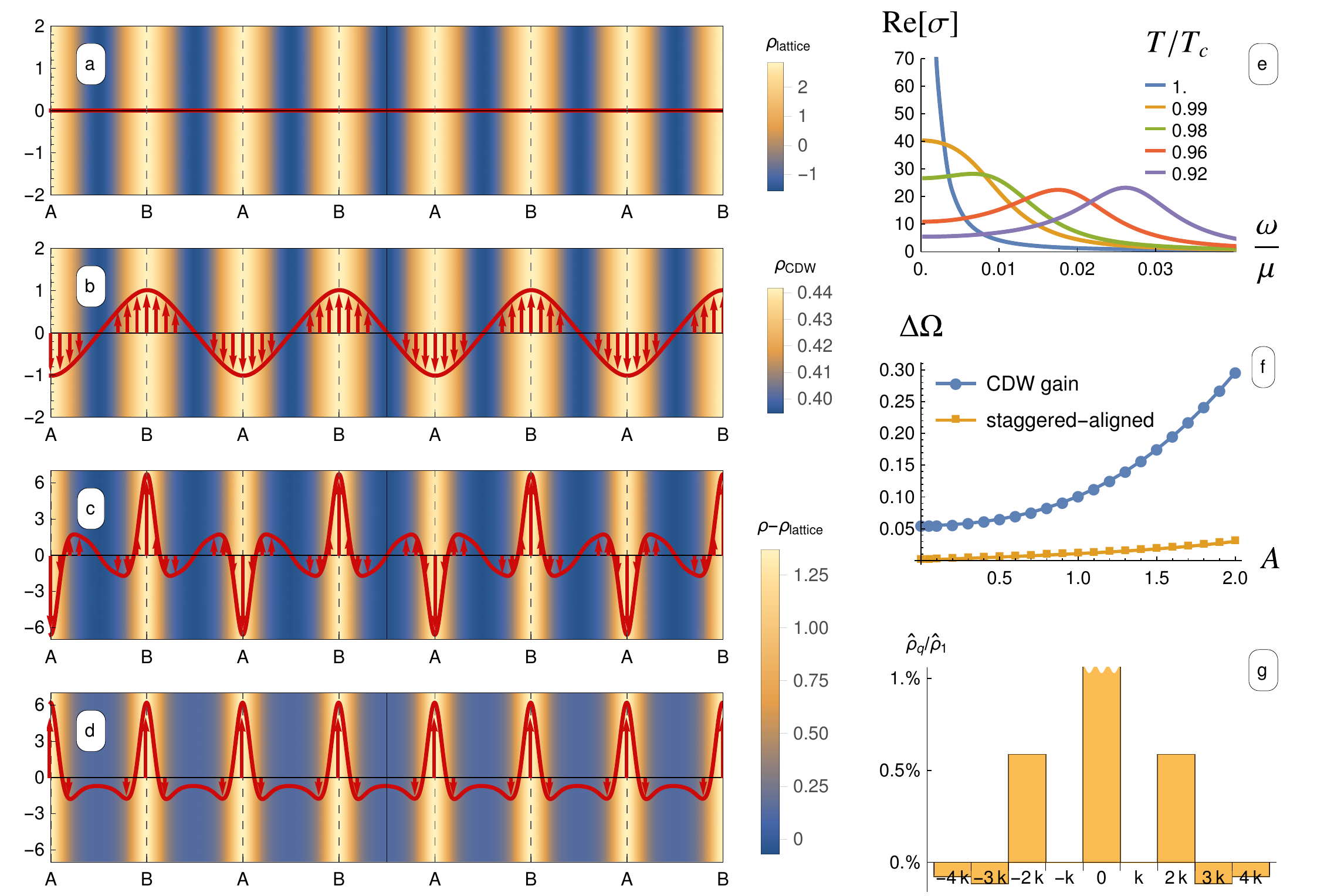}
 \caption{\small \textbf{Formation of holographic Mott insulator} Left panel:
  Profiles of the spontaneous 
 currents (arrows) and charge density
  (colour) in the (a) ionic lattice without spontaneous order
  (unbroken phase); (b) purely spontaneous intertwined CDW state, (c)
  commensurately locked Mott state and (d) State with aligned
  currents. Due to intertwinement of order this state has
  a different charge density than the Mott state. Note that the total current is zero in both staggered and aligned states. All solutions are
  at a fixed chemical potential with $T=0.01 \mu$, lattice
  potential strength  $A=2$, and $\theta$-coupling $c_1=17$. \\
Right panel (e): \textbf{Evolution of optical conductivity} upon the
phase transition from metallic to Mott state. A sharp Drude peak is seen
in metallic state which is pinned and broadened after the phase
transition. (f): \textbf{Energy scales and superexchange}: The grand
thermodynamic potential difference between unbroken phase and Mott
state (blue line) and for the Mott state and aligned currents state
(yellow line), as a function of the strength of the lattice
potential. Clearly, the energy scale of the current ordering lacks
behind the one of the charge ordering when the lattice becomes
strong. Note that the grand canonical ensemble is required due to the charge difference between the two current
configurations. 
(g): \textbf{Higher harmonics}: The difference between Fourier transform of the charge density $\hat{\rho}_q$ of the aligned (d) and staggered (c) states. Both spectra are normalized with respect to the lattice periodic mode ($\hat{\rho}_k$). The enhancement of $2k$ mode is seen for the aligned state, showing that it has twice the number of CDW per unit cell. The enhancement of the homogeneous component by $\sim10\%$ in not shown.
}
\end{figure}

\begin{figure}[t!]
\centering
\includegraphics[width=1.\linewidth]{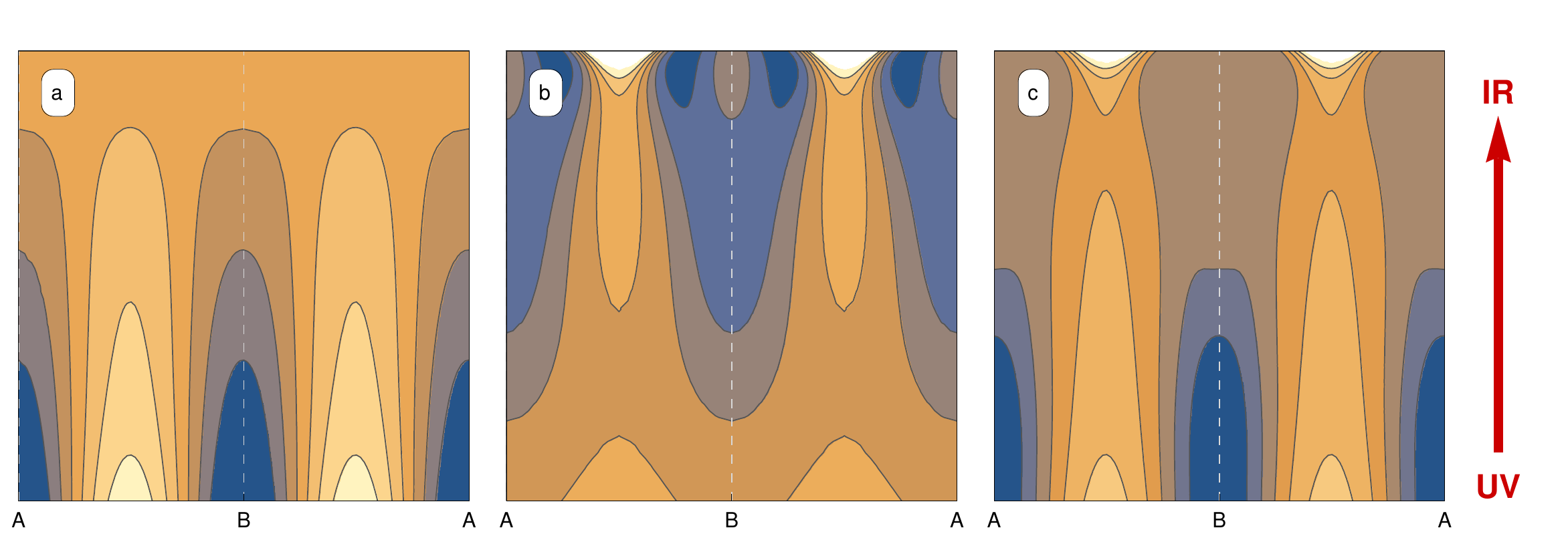}
\caption{\label{fig:bulk_profiles} \textbf{Holographic RG pattern}:
  the profiles of the electric field strength $\partial_z A_t$ (see eq. \eqref{A ansatz} in the Supplementary Material) in the
  gravitational theory encoding the RG flow from the UV (bottom) to
  the IR (top) of (a) the pure lattice that is sourced in UV and decreases to irrelevancy in the IR (b) The spontaneous charge density wave that emerges in the IR, but don't have sources in UV (c) The lock-in that forms the Mott state. The CDW, locks to the lattice at intermediate scales and introduces the relevant explicit translation symmetry breaking in IR, giving rise to the insulating state.}
\end{figure}

One can also introduce a background periodic potential that breaks
translational symmetry explicitly by representing the ion lattice in
terms of a
spatially modulated chemical potential in the field theory
\cite{Flauger:2010tv,Liu:2012tr,Horowitz:2012ky,Horowitz:2012gs,Donos:2014yya,Rangamani:2015hka}.  It is easy to incorporate complicated forms of such ``pseudo potentials'' but we will focus here on the
simplest choice in the form of a uni-directional single harmonic
potential with wave vector $k$ and relative amplitude $A$: $\mu(x) =
\mu_0 (1 + A \cos(k x))$. (Fig.\,1a)

Combining these two allows us to study 
spontaneous holographic crystallisation in the presence of a
background lattice. The crystal tends to form with a preferred
intrinsic wave-length $p_0$ set by the parameter $\vartheta$ and the
scale of the mean chemical potential $\mu_0$. In the presence of a periodic potential characterized by wave vector $k$, 
one anticipates that one is considering incommensurate systems studied thoroughly in the past in classical matter \cite{pokrovsky1979ground,bak1982commensurate}. As already anticipated in \cite{Andrade:2017leb}, when these periodicities 
are sufficiently close together one expects a ``commensurate lock-in'' of the spontaneous crystal, gaining additional stability.  {\em These lowest order commensurate states are the holographic incarnations of Mott insulators} 
(Fig.\,1c). How literal is this ``assignment''? It is instructive to consider first the optical conductivity. In the absence of the periodic potential one finds a ``diamagnetic'' delta function peak at zero frequency 
at all temperatures. The reason is that every finite density system is
a perfect metal in the spatial continuum limit since total
momentum is conserved. The formation of a crystal at $T_c$
  spontaneously breaks translational invariance, and a longitudinal phonon emerges    
that is dual to a perfect current: the sliding mode. When we now
switch on a sufficiently strong explicit commensurate background
potential, this sliding mode will acquire a {\em mass} since the
crystal gets pinned and it can no longer freely slide. This
  reveals itself in the optical conductivity
  (Fig.\,1e). As the crystal forms below $T_c$, the
  metallic Drude peak rapidly moves off to finite frequency
  corresponding to the pinning of the sliding mode. The mode itself
  broadens first due to increased translational symmetry
  breaking from the crystal. The resulting optical conductivity at
  $T<T_c$ strongly resembles that of hard MI with a broadening due to Hubbard interband transitions. 

From Fig.\,1c one infers that the background lattice
enhances the spontaneous order. This can be further quantified by
computing the energy difference between the crystallised and
uncrystallised phase of the 
strange metal which increases steeply as function of the lattice
amplitude (Fig.\,1f). Visually one notices
that the  currents are generically enhanced in the regions where the spontaneous 
charge density wave has a maximum and the current density is
effectively localized in these regions. This charge localization together
with the alternating pattern of
these currents immediately calls to mind the hard antiferromagnetic
Mott insulator with staggered spins.

This suggests that other current patterns also exist. As we shall
show, bulk saddle points (local minima in the grand thermodynamic potential $\Omega$) 
exist where the currents are aligned. As function of the lattice potential, the energy difference between these two 
configurations is much smaller than the energy difference between the
CDW ordered and the uncrystallized state. 
This implies that in the presence of a large lattice potential,  current-current dynamics is governed by a different
scale than charge dynamics. This is in analogy with the
spin-charge separation in conventional Mott insulators, where below
the Mott transition one is dealing with an spin only system, with the {spin} (dis)order governed by
effective ``super exchange'' interactions which are much smaller than the scale associated with the Mott-insulator itself. 
The ramification is that the dynamics we visualize in the holographic model is not what actually would happen in the true physical system. For strong lattice potentials, one first encounters the onset of the charge density wave order at the transition temperature. Only at lower temperature will the additional staggered current symmetry breaking occur, since the latter will remain thermally disordered at temperatures larger than the current-current ”exchange” parameter. A full computation in the holographic model with no saddlepoint approximation, will exhibit this physics. 
The solutions here focus on a single saddlepoint only, which is of the
Mott insulator type when the wave-vectors of the spontaneous and ionic
crystals are close to each other.

A highlight of holography is that the extra dimension 
of the gravitational theory can be interpreted as the ``scaling direction'' of the renormalisation group of the dual field theory with the UV fixed point located on the boundary of AdS.  
This  yields a vivid renormalisation group view on the way that the Mott-insulator is formed. 
The irrelevancy of the explicit potential is illustrated in Fig.\,2a: the electric field sourced by the external potential falls off moving from the boundary to the deep interior. 
The spontaneous crystal displays however precisely the opposite flow: it is relevant in the IR without having any sources in the UV (Fig.\,2b).
One can now read off the commensurate pinning mechanism from the ``scaling diagram'', Fig.\,2c: this pinning occurs at intermediate energy scales. 
One sees that ``halfway'' the radial direction the (decreasing) external potential starts to overlap with the (increasing) ``hair'' coding for the spontaneous crystal.

\bigskip

The important question with reference to the cuprates is: what
happens when these holographic Mott insulators are doped? Above we
  tuned the wavevectors of the explicit lattice and spontaneous
  crystal to be the same. Adding charge the crystal will want
to form at a different
intrinsic wave vector $p_0$ but the external lattice potential may force it to
acquire an altered one ($p$).
Driving $p$ away  from $p_0$ will, however, cost potential energy due to the elastic response of the crystal, so the resulting value of $p$ is determined dynamically by these two competing mechanisms.
This is a motive familiar from the study of classical incommensurate
systems \cite{bak1982commensurate,FKbook} and one anticipates that generically  this will promote values of $\frac{p}{k}$, which are the rationals of small coprime integers:
 these are the \textit{higher order commensurate points}. The states labeled by the different fractions 
$\frac{p}{k}$ form a set of the local minima in the thermodynamic potential $\Omega$ and the true ground state corresponds to the one with minimal $\Omega$. We performed extensive numerical computations 
spanning a large parameter space to identify these saddle points.  

The lowest order commensurate state $p/k = 1/1$ is obviously the Mott insulator we just discussed in detail. In analogy with the conventional picture of adding microscopic charges per unit cell, we prescribe the doping level as the excess charge per lattice period compared to the Mott insulating state. We normalize by assigning doping level $100\%$ to the $p/k = 2/1$ state, which has exactly one additional period of spontaneous CDW per unit cell (see Supplementary Material).
In practice, adding excess charge to the system is  accomplished by adjusting the chemical potential, while keeping the lattice wave vector fixed.
 The result is 
summarized in Fig.\,3a. The noteworthy
  aspect is that, due to the lock-in, some commensurate points stay
stable for a range of dopings, displaying a ``Devil staircase'' like
behavior  familiar from classical incommensurate systems. We
  shall return to this shortly.

\begin{figure}[!ht]
\center
\includegraphics[width=1\linewidth]{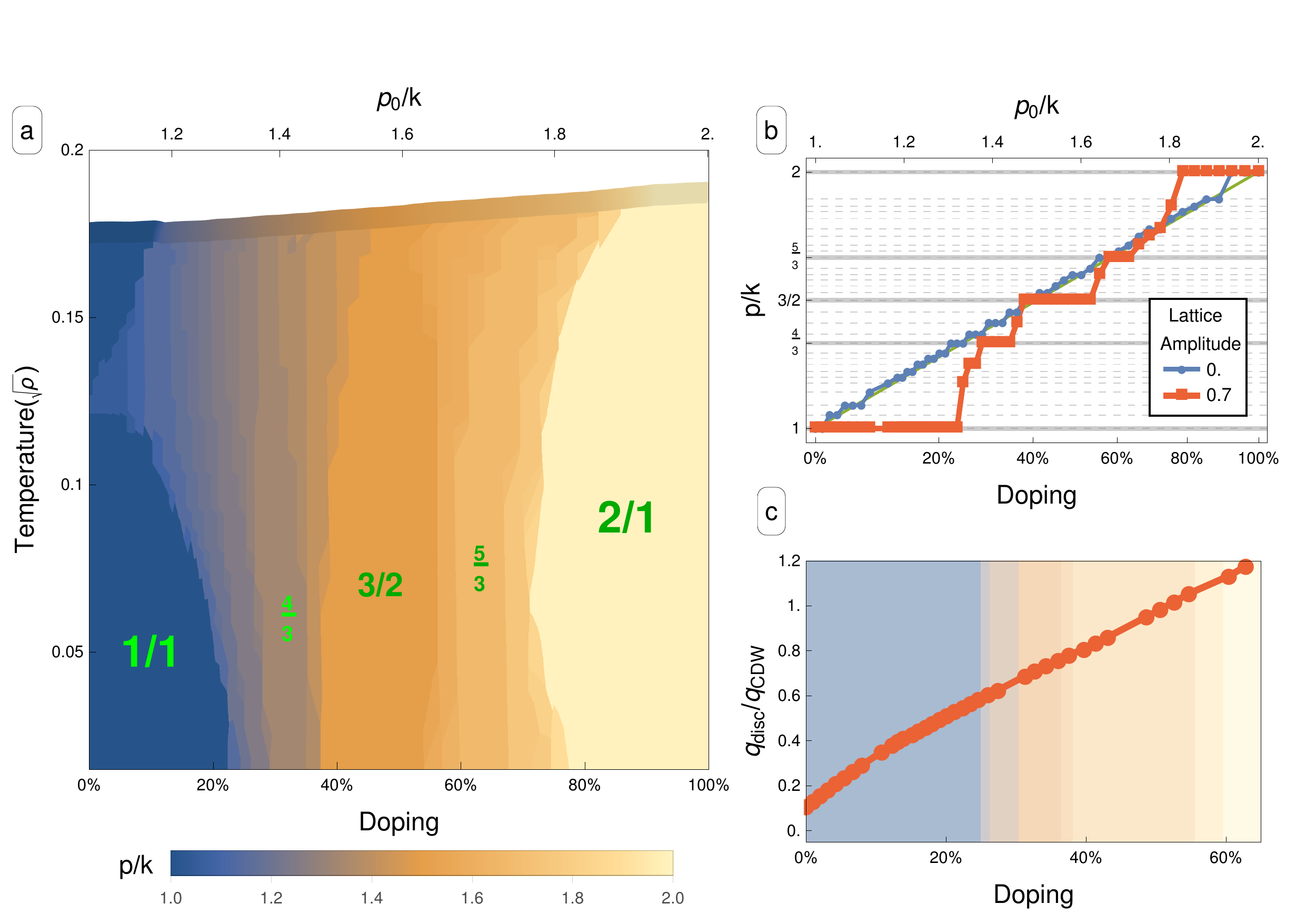}
\caption{\small (a) \textbf{The commensurate/incommensurate phase diagram.}
  Colour shows the thermodynamically preferred commensurate fraction
  as a function of doping and temperature. The regions of stability of
  the leading $1/1$, $2/1$, and, as temperature is lowered, higher
  $3/2$, $4/3$, $5/3$ commensurate points are seen.  The shaded line
  on top shows the result of perturbative analysis of
  instabilities (see Supplementary Material). Data is taken for
  $A=0.7, c_1=17$  
(b) \textbf{The commensurate plateaux} as seen on the
  fixed temperature cut ($T\approx0.01 \sqrt{\rho}$) of the phase
  diagram. Importantly, the commensurate states stay stable for
  a range of the charge density values. Higher commensurate points
  $3/2$ and $4/3$ correspond to $2a$ and $3a$ discommensuration
  lattices, respectively ($a$ -- the lattice constant). $p_0/k$ shows the relation between spontaneous momentum of the free CDW versus the momentum of the lattice. Blue points represent the result in absence
  of commensurate lock in: when the amplitude of the lattice potential
  vanishes the preferred momentum of the structure equals the spontaneous one. The gridlines show the mesh of numerical study, where different saddle points were obtained. 
(c) \textbf{Charge of a discommensuration} as a function of doping
(red line), measured in units of CDW charge density integrated over a unit cell (see
\eqref{equ:qCDW}). The charge is obtained by considering an isolated discommensuration (one over 19 unit cells) and subtracting the contribution of the parent Mott state. The charge changes continuously as the doping is
increased. Shading shows the preferred commensurate fraction as on (a), no signs of plateaux are seen in the charge of discommensuration.}
\end{figure}

\begin{figure}[!ht]
\center
\includegraphics[width=0.8\linewidth]{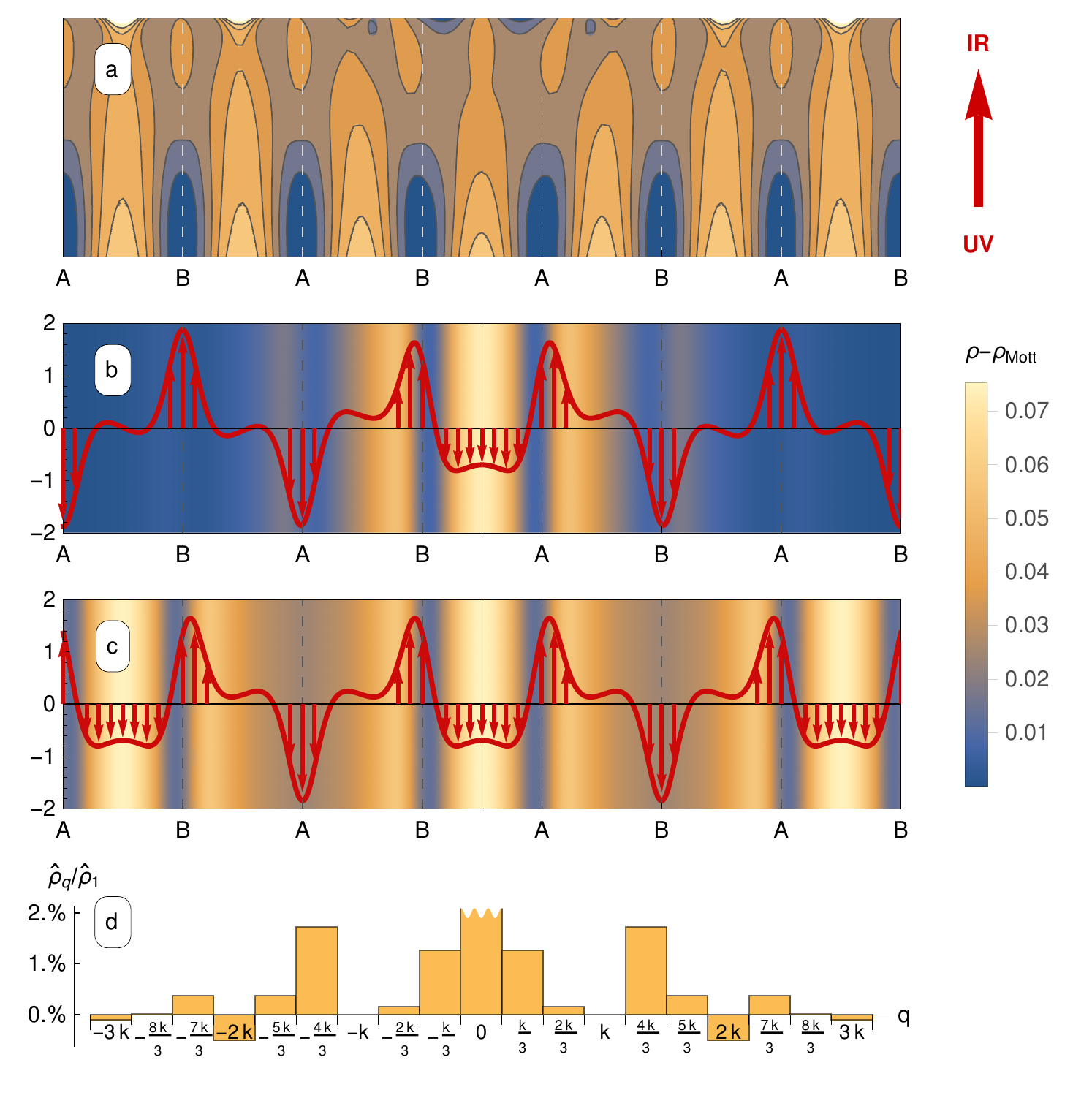}
\caption{\small
(a) {\bf RG structure of an isolated discommensuration} (same notation
as on Fig. 2) The UV lattice lock in the IR structure everywhere outside the core of a discommensuration. In the core the ``dislocation'' is seen in the electric field, accounting for an excess period of the spontaneous structure in IR.
{\bf Profiles of the discommensurations.} Currents (arrows) and
    charge density (color) are shown for {\bf (b) an isolated discommensuration}. The domain wall in the staggered current (defined using A and B sublattices) is clearly seen as well as excess charge in the core of a discommensuration. {\bf(c) Higher commensurate state} with $p/k = 4/3$. The state displays superstructure  with a period of 3 unit cells -- 3a discommensuration lattice.
    The charge profile is normalized with respect to the corresponding Mott state with $A=0.7, T=0.01 \mu$.
    (d) {\bf Spectrum of discommensuration lattice}. The difference between the Fourier modes of the lattice and the parent Mott insulator. Both spectra are normalized with respect to the lattice periodic mode $\hat{q}_k$. The fractional Fourier modes are clearly enhanced in the disc. lattice state. The homogeneous mode reaches $\approx 11\%$ and is not shown. 
}
\end{figure}

Let us first discuss the structure of these higher order commensurate states as formed at low temperatures in sufficiently strong background potentials. The periodicity mismatch 
(the deviation of $p/k$ from $1/1$)  is concentrated in
localised solitonic textures, see Fig.\,4b -- the \textit{discommensurations} familiar from classical incommensurate systems (see also Supplementary Material). 
This is  not completely surprising since
discommensurations are rather ubiquitous when dealing with incommensurate
systems. It is entertaining to observe  how the discommensurations follow the
renormalization group in the extra dimension of the gravity system
(Fig. 4a). The UV lattice almost everywhere locks-in the IR charge
density wave, except at the 
discommensuration core where a curious dislocation is formed in the
electrical flux in the radial direction of the gravitational theory. 
 
The noteworthy aspect is that there is additional structure: these discommensurations  are at the same 
time \textit{domain walls} in the staggered current order (Fig.\,4b). Considering the current order as being analogous to the antiferromagnetic spin systems found in the standard (doped) Mott-insulators, these are just like 
the famous ``stripes'' observed in the $La_2CuO_4$ (214) family of high
Tc superconductors \cite{tranquada1995evidence} and in other doped
Mott-insulators \cite{vojta2009lattice}: in the cuprate stripes the doped charge accumulates at the spin-pattern 
domain walls and the same is happening here
(Fig.\,4b). These stripes were
actually discovered theoretically on basis of Hartree-Fock
calculations  well before the experimental observation
\cite{zaanen1989charged}. It is therefore clear that  Hartree-Fock form of ``intertwinement'' of spin and charge
order originates in the  quantum mechanics of localized
  electrons.  The product state nature of these
mean-field stripes is revealed by the rule that the stripes have a
preferred density (typically, one hole per domain wall unit
cell). This is crucially different here: {\em the ``holographic stripes'' have no preferred charge density.}

In Fig.\,3b we highlight that the higher order commensurate
plateaux are in fact stable across a range of dopings. This has the
implication that the charge density inside the ``stripes'' continuously
varies over a considerable range
(Fig.\,3c). This is a natural outcome of the
absence of localized quantized charge in this
entangled matter. 

This observation is directly relevant to
experiment.  It has proven difficult to explain the dependence of the
charge order wave vector on doping  as observed in the various
cuprates in terms of conventional ``product state'' charge
  density waves. In most cuprates it appears that the periodicity locks
locally at 4 lattice constants in the whole doping range
\cite{comin2016resonant,fradkin2015colloquium,keimer2015quantum,mesaros2016commensurate}
(the exception are 214 stripes which do show a sense of
preferred charge density in a limited doping range). Product
  state, localized electron-based Mott insulator models cannot do
this.
 This is supported by a
puzzling result obtained
recently in the context of 
numerical approaches to the doped Hubbard model
\cite{zheng2016stripe,huang2016numerical}. This revealed that stripes
are ubiquitous, with a similar surprise that these lack a preferred charge
density. These heavy numerical methods wire in entanglement,  the same generic motive that is hard wired in
holography. The locked-in periodicity of stripes can therefore
be seen as another
compelling indication that strong entanglement underlies cuprate strange metals.   We present it as
a challenge to both theorists and experimenters to see whether explicit signatures can be found 
demonstrating that this behaviour is indeed caused by  many-body entanglement as suggested by holography.

\begin{figure}[ht]
\center
\includegraphics[width=0.9 \linewidth]{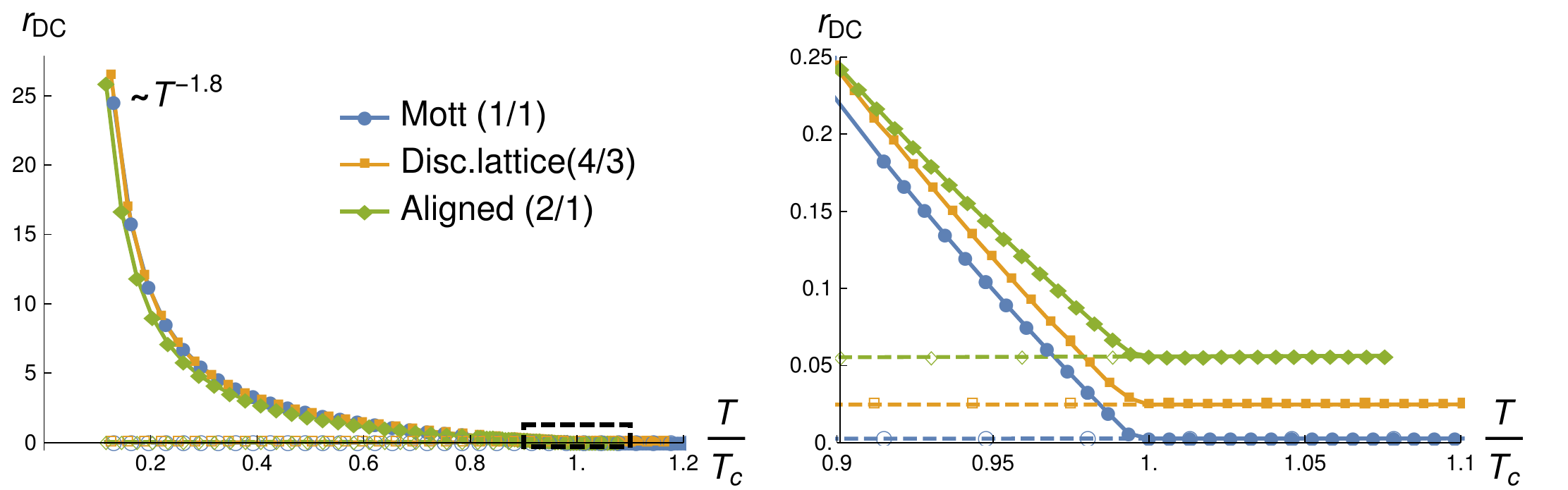}

\caption{\textbf{DC resistivity of the holographic Mott insulator.} Left panel: Different commensurate states share similar behavior, revealing power law scaling at low temperature. This signals the presence of quantum critical degrees of freedom, which remain ungapped. Right panel: At the critical point no abrupt change of resistivity is seen, suggesting a metal-insulator crossover. The dashed lines show resistivity in the unbroken state without CDW, which is metallic. Temperature is scaled with respect to the corresponding critical temperature of a given state (see e.g. Fig.\,3a)}
\end{figure}

Last but not least we should clarify how the ``insulating'' nature of
  the Mott state and striped phases shows up in the
DC resistivity.
In Fig.\,5a we 
show the results for a variety of cases including the $1/1$ ``Mott
insulator''.   The
surprise is that in all cases
at temperatures well below $T_c$ the resistivity diverges {\em
  algebraically} instead of exponentially, approximately as $r_{DC}
\sim T^{-1.8}$. This is suggestively similar to the slow increase of
the resistivity observed in the spin-stripe ordered cuprates
\cite{boebinger1996insulator,laliberte2016origin}. It represents a
long standing experimental puzzle, but it could well be just  a version of the algebraic behaviour we obtain from holography. 
 
Theoretically, it is a
consequence of one of holography's big mysteries.  Closer inspection reveals that the transport associated with the pinned charge density wave itself is gapped at low temperatures, in the guise of the identification of the pinned crystal with the Mott insulator.  However, for reasons associated with the universality 
  of black holes, holography invariably
 predicts that {\em inside} the low temperature symmetry broken states
 a conducting quantum critical infrared persists \cite{Grozdanov:2015qia} although with scaling properties 
 different from those of the high temperature strange metal \cite{Donos:2014oha}.  In the presence of  translational symmetry breaking the scaling properties of the electrical current operators are typically modified, 
 causing the algebraic rise of the resistivity in Fig.\,5a.

\bigskip 

Summarizing,  
we have identified the holographic analogue of Mott insulators having crucial 
properties in common with the conventional variety, such as a Mott gap
and the mechanism of super exchange interactions. By construction this analogue displays intertwinement of charge order with spontaneous currents and parity breaking of the continuum. Doping this state results in textures that have striking similarities with the stripe phases 
found in cuprates. \\ 

The distinct novel ingredient is that holographic matter should be strongly entangled. 
This new principle is reflected in specific predictions for unconventional properties due to the ``unparticle'' nature of this matter
that beg to be further investigated in the laboratory.
Firstly, when electron matter is strongly entangled the quantisation of the electrical charge 
will be washed out, with the ramification that there should be no ``quantised'' relationship between 
ordering wave vectors and the number of electrons, either in momentum or real space. This may 
explain the doping-independence of the charge order in underdoped cuprates. 
Secondly, we find a restructuring of the quantum critical infrared scaling in the ordered states, 
reflected in a low temperature algebraic behavior in transport. 
This is not at all understood theoretically. 
One interpretation is that spontaneous symmetry breaking from the 
highly entangled strange metal states does not result in a purely ``classical'' product state, so it does not gap out all of the 
degrees of freedom.  Instead, there is now a sector of strongly entangled quantum critical states with modified properties. This is what holography predicts. 
  However, holography originates critically in a matrix large 
$N$ nature of the microscopic degrees of freedom in the field  theory \cite{Zaanen:2015oix}, lacking any relation with the electrons of condensed matter physics.
One relies on the universality of the RG that the IR physics has
  a larger applicability than matrix large $N$ field theories, and
  captures strongly entangled physics that may
apply here.
Is this at work in underdoped cuprates? Our suggestive findings here
  together with other experimental indications, are a clear call to nail this down by a focused experimental
effort.




{\bf Acknowledgments} \\
We thank 
Jerome Gauntlett, Aristomenis Donos, Blaise Gouteraux, Nikolaos Kaplis, Christiana Pantelidou, Jorge Santos for insightful discussions. 

The research of K.S., A.K. and J.Z. was supported in part by a VICI (KS) award of the Netherlands Organization for Scientific Research (NWO), 
by the Netherlands Organization for Scientific Research/Ministry of Science and Education (NWO/OCW), and by the 
Foundation for Research into Fundamental Matter (FOM).
The work of T.A. was supported by the European Research Council under the European Union's 
Seventh Framework Programme (ERC Grant agreement 307955) and in part by the Newton-Picarte Grant 20140053. 

Numerical calculations have been performed on the Maris Cluster of Lorentz Institute


\clearpage

\appendix
\section{Supplementary material}

\subsection{The holographic setup} 
\label{sec:the_holographic_setup}

We consider the model of \cite{Donos:2011bh} which consists of 3+1
dimensional Einstein-Maxwell theory coupled to 
a neutral pseudo scalar. Following the conventions of \cite{Withers:2013kva, Withers:2013loa}, we write the action as 
\begin{equation}\label{S_full}
	S = \int d^4 x \sqrt{- g} \left( R - \frac{1}{2} (\partial \psi)^2 - \frac{\tau(\psi)}{4} F^2 - V(\psi) \right)
	 - \frac{1}{2} \int {\vartheta}(\psi) F \wedge F + S_{bndy}
\end{equation}
\noindent where 
\begin{equation}
\label{Sbndy}
	S_{bndy} = - \int d^3 x \sqrt{- h} (K - 4 + \psi^2)
\end{equation}
Here $F=d{\cal A}$ is the field strength associated to the Maxwell field ${\cal A}$, while 
$h$ is the metric induced at the boundary with extrinsic curvature $K$.
We have checked that the boundary term \eqref{Sbndy} obtained in \cite{Withers:2013loa} suffices to properly renormalize the action \cite{deHaro:2000vlm}. 
The AdS/CFT dictionary \cite{Maldacena:1997re,Witten:1998qj,Gubser:1998bc} relates 
the boundary asymptotics of the fields in \eqref{S_full} to the sources and responses 
of the energy-momentum tensor, electromagnetic currents and parity-odd
order parameter in the dual 2+1 dimensional field theory.
Following 
\cite{Donos:2011bh, Withers:2013kva, Withers:2013loa, Donos:2013wia}, we choose the couplings as 
\begin{equation}
\label{equ:potentials}
	V(\psi) \equiv 2 \Lambda +W(\psi) = - 6 \cosh (\psi /\sqrt{3}) , \quad \tau(\psi) = \frac{1}{{\rm cosh} (\sqrt{3} \psi)}, \quad \vartheta(\psi) = \frac{c_1}{6 \sqrt{2}} \tanh(\sqrt{3} \psi),
\end{equation}
This model is bottom-up, but similar couplings can be obtained from
dimensional reduction of supergravity \cite{Gauntlett:2009bh}.  
Note that in these conventions the cosmological constant is $\Lambda = - 3$ and the mass of the scalar is $m^2 = -2$. 
The equations of motion admit the translational invariant
Reissner-Nordstr\"om (RN) charged black hole solution which can be written as
\begin{equation}\label{RN soln}
	ds^2 = \frac{1}{z^2} \left( - f(z) dt^2 + \frac{dz^2}{f(z)} + dx^2 + dy^2 \right) , \quad A = \bar \mu (1 - z) dt , \qquad \psi = 0
\end{equation}
\noindent where
\begin{equation}\label{f RN}
	f = (1-z)\left( 1 + z + z^2 - \bar \mu^2 z^3 /4 \right)
\end{equation}
In these coordinates the boundary is located at $z=0$ while the horizon is at $z=1$. The chemical potential in the 
dual theory is given by the constant $\bar \mu$.
The Hawking temperature of this black hole reads
\begin{equation}\label{eq:T}
 	\boldsymbol{\mathit{T}} = \frac{12 - \bar \mu^2}{16 \pi}.
\end{equation} 
\noindent 

We will be interested in stationary time-independent configurations of the form 
\begin{align}\label{ds2 anstaz}
	ds^2 &= \frac{1}{z^2}\left( - Q_{tt} f(z) dt^2 + Q_{zz} \frac{dz^2}{f(z)} + Q_{xx} (dx + Q_{zx} dz)^2 + Q_{yy} ( dy + Q_{ty} dt )^2  \right), \\
\label{A ansatz}
	{\cal A} &= A_t dt + A_y dy 
\end{align}
Here, all unknowns
are functions of the holographic coordinate $z$ and the boundary coordinate $x$.  
We search for black holes with a uniform spatially constant
temperature, which means that in the near horizon all functions must
be regular except $f(z)$. With this assumption, the equations of motion require $Q_{tt}(1,x) = Q_{zz}(1,x)$, which in turn implies that the surface gravity is constant and given by \eqref{eq:T}, see e.g. \cite{Horowitz:2012ky}. 
Since we are interested in a dual field theory living in flat space, we require the metric to be asymptotically AdS as $z \rightarrow 0$. 
The UV boundary conditions on the gauge field will be dictated by our choice of explicit or spontaneous breaking of translations. 
Following the standard AdS/CFT prescription, we relate the boundary data to the dual field theory one-point functions. In particular, 
from the UV expansions
\begin{align}
\label{series UV 1}
	Q_{tt} & = 1 + z^2 Q_{tt}^{(2)}(x) + z^3 Q_{tt}^{(3)}(x) + O(z^4) \\
  A_t &= \mu(x) - z \rho(x) + O(z^2) \\
  A_y & = z J_y(x) + O(z^2)  \\
\label{series UV 4}
\psi &=  z^2 \psi^{(2)}(x) + O(z^3) 
\end{align}
we obtain that the coefficients $\mu(x)$, $\rho(x)$, $J_y(x)$ and $\psi^{(2)}(x)$  determine the chemical potential, charge density, current density and 
pseudoscalar parity breaking order parameter of the dual theory, $Q_{tt}^{(2)}$ is a function of the sources only and the energy density is given by 
\begin{align}
	 \epsilon(x) &= 2 + \frac{\bar \mu^2}{2} - 3 Q_{tt}^{(3)}(x)  
\end{align}

In order to break translations explicitly, we follow \cite{Flauger:2010tv,Horowitz:2012gs} and introduce a spatially modulated chemical potential, fixing $A_t(z=0,x) = \mu(x)$ with 
\begin{equation}\label{eq:mu x}
	\mu(x) = \mu_0 (1 + A \cos(\boldsymbol{\mathit{k}} x) )
\end{equation}
Without loss of generality, we set $\mu_0 = \bar \mu$ \cite{Rozali:2012es}. Unless otherwise stated, we 
express the dimensionful parameters of the model, denoted up until now in bold script, in units of $\bar \mu$ by making the replacements
\begin{equation}
\label{mu units}	
		\boldsymbol{\mathit{T}} = T \bar \mu, \qquad  \boldsymbol{\mathit{k}} = k \bar \mu, \qquad 
		\boldsymbol{\mathit{p}} = p \bar \mu
\end{equation}
When $\psi = Q_{ty} = A_y = 0$, all profiles acquire modulation along
$x$ solely due to the $x$-dependent boundary conditions, so these
solutions  represent states which break translations explicitly only. They were termed ``ionic lattices'' in \cite{Horowitz:2012gs}. As shown there, 
for small $\omega$, the optical conductivity can be approximated by a Drude peak with finite DC value. 
Moreover, these lattices are irrelevant in the IR, in the sense that the near-horizon geometry approaches the translationally invariant 
charged black hole solution
as we lower the temperature \cite{Donos:2014yya}. Because of this, one
can think of the lattice as an UV-based structure (see
Fig.\,\ref{fig:bulk_profiles}).\footnote{Holographic models exist
  where the lattice stays relevant in the IR \cite{Rozali:2012es}. We will not consider
  those here.}

In order for translations to be broken only spontaneously, the boundary conditions need to be translational invariant in the UV, 
so we take the chemical potential to be constant $A= 0$ in \eqref{eq:mu x}, along with the vanishing of the sources dual to the 
leading terms in $A_y$ and $\psi$, as reflected by \eqref{series UV
  1}-\eqref{series UV 4}; see also \cite{Withers:2013loa,
  Withers:2013kva, Donos:2013wia, Rozali:2012es}. The spontaneous
breaking of translations --- an IR effect ---is due to  near-horizon
instabilities induced by the topological $\vartheta$-term in
\eqref{S_full}. Accordingly, the effect of the spontaneous breaking localizes near the horizon of the black hole (see Fig.\,\ref{fig:bulk_profiles}). 
The features of the resulting spontaneous structure
are strongly dependent on the chemical potential $\bar \mu \neq 0$ and
the value of the coupling $c_1$ in \eqref{equ:potentials}: we observe
that increasing $c_1$ rises the critical temperature and makes the
spontaneous crystal more stable, leading to more pronounced
commensurate effects, which we are after. For this reason we choose $c_1=17$, as opposed to \cite{Donos:2011bh, Withers:2013kva, Withers:2013loa, Donos:2013wia}.

The arising spontaneous structure is characterized by the oscillating
values of $A_y$ and $\psi$, which results in the observable staggered
currents $J_y(x)$ in the dual theory. At the nonlinear level the $A_t$
component of the gauge field also becomes modulated with {\em twice} the momentum of $A_y$ or $\psi$ due to the quadratic interaction in $\vartheta$-term \eqref{S_0}, see Fig.\,8b. 
The modulation of $A_t$ corresponds to the formation of a charge density wave (CDW) on the boundary, which we write as
\begin{equation}
\label{equ:rhoCDW}
\rho_{CDW} = \rho_0 + \delta \rho \cos(\boldsymbol{\mathit{p}} x).
\end{equation}
We use the momentum of this CDW $\boldsymbol{\mathit{p}}$ as the
defining momentum to describe the spontaneous structure, i.e. the
staggered currents have momentum $\boldsymbol{\mathit{p}}/2$. This notation
is {\em different} from the previous studies \cite{Donos:2011bh, Withers:2013kva, Withers:2013loa, Donos:2013wia, Andrade:2017leb}. In order to compare the results one has to make a redefinition $p_{here} = 2 p_{there}$.

\bigskip
 
It is instructive to first consider spontaneous symmetry breaking
perturbatively by taking $\psi$, $A_y$ and $Q_{ty}$ to be small
functions in a Fourier basis of momentum $\boldsymbol{\mathit{p}}$ and
zero frequency and linearize. Linear instabilities exist for $T < T_c$ and arrange themselves in a ``bell-shaped'' curve in the $(T,p)$ plane \cite{Donos:2011bh}, much as in the 5-dimensional case first discussed in \cite{Nakamura:2009tf}, see Fig.\,6a. 
The critical temperature corresponds to the maximum of the 
bell $T_c^{RN} =  0.147$, attained at a critical momentum $p_c^{RN} = 1.33$.  
Inside the bell-curve at any given temperature below $T_c$, one can construct non-linear solutions for a range of values of $p$. However, the ones that minimize the (spatially averaged) thermodynamic potential play a special role since they are the ones which dominate the thermodynamic ensemble. 
These thermodynamically preferred solutions for all $T$ form a line $p_0(T)$ inside the bell-curve, which in general deviates from $p_c^{RN}$.

\begin{figure}[ht]
\center

\includegraphics[width=0.9\linewidth]{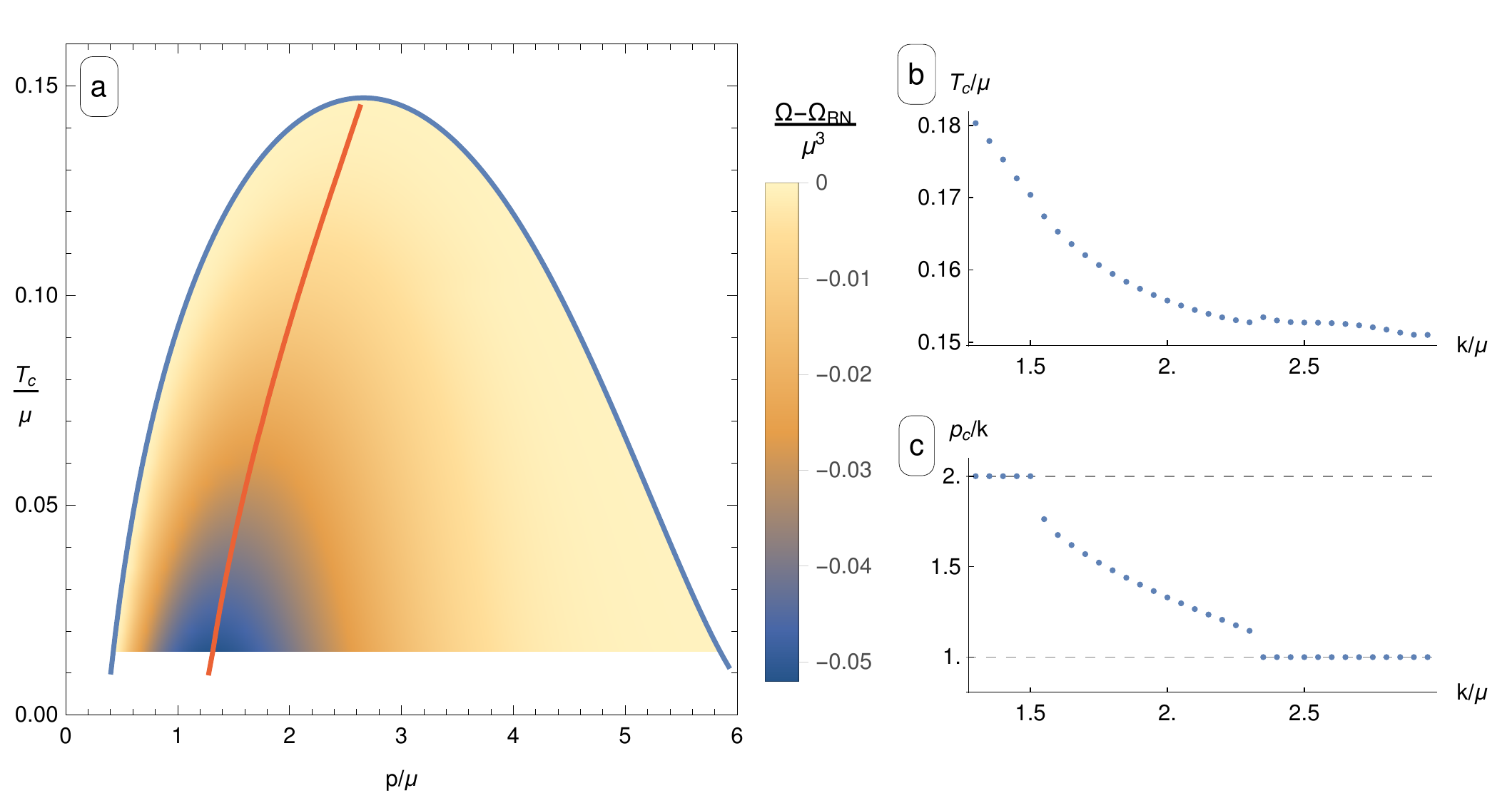}

\caption{\label{fig:perturb} Left panel: {\bf Perturbative analysis of spontaneous
    CDW formation:} \textbf{(a)}Blue curve: Unstable modes of the RN
  solution. Orange curve: the thermodynamically preferred momentum
  value $p_c(T)$ of the nonlinear solutions. \\
Right panel: {\bf Pertubative lock-in of the spontaneous CDW with an
  explicit lattice:} \textbf{(b):} Behaviour of the critical temperature for varying lattice
  momentum $k$. For large $k$, we approach the translationally invariant value $T_c^{RN} \sim 0.15$.} \textbf{(c)} Unstable modes
  in the presence of potential of amplitude $A=0.7$ for varying lattice
  momentum $k$. We observe the plateaux at $p_c/k = 1$ and $p_c/k =
  2$. 
\end{figure}

The aim of this work is to study the interplay of the explicit and spontaneous symmetry breaking phenomena. We can do it in two ways: we can start with an ionic lattice and observe how the instabilities towards the formation of spontaneous structures develop, or begin with a configuration that breaks the translations of the RN solution spontaneously and introduce a modulated source as in \eqref{eq:mu x}.
Following the former procedure, we need to examine the unstable modes of the pure ionic lattice solution. 
The study of these unstable modes was undertaken in \cite{Andrade:2017leb}, which revealed an interesting lock-in pattern of the spontaneous to the explicit structure.
We reproduced the calculations of \cite{Andrade:2017leb} for the
parameters which will be used in our nonlinear study: $A = 0.7$
and $c_1=17$, see Fig.\,6b,c. We observe the lock-in of the spontaneous structure indicated by the plateaux at $p_c/k = 1$ and
$p_c/k = 2$. Importantly, the higher order commensurate fractions,
i.e. $p_c/k=3/2$, cannot be observed in the perturbative approach. In the regime where the spontaneous structure is infinitesimal
the total solution including the perturbative modes is forced by the
lattice to be periodic with
momentum $k$. Hence near critical
temperature all possible commensurate fractions of $p_c/k$ are
integers. This changes as soon as one considers finite amplitude of
the spontaneous structure in fully nonlinear approach; the result is
Fig.\,3a in the main text.

\subsection{Full backreacted solutions} 
\label{sec:nonlin_solution}

We construct the fully backreacted nonlinear solutions by observing how a given purely spontaneous structure which arose from RN gets modified as we place it on top of an ionic lattice potential by tuning its amplitude from $A = 0$ to a finite value. 

We wish to study the thermodynamic stability of the so constructed configurations, by finding the ones that minimize the spatially averaged thermodynamic potential
\begin{equation}\label{eq:free energy}
	 \Omega(x) = \epsilon(x) - \boldsymbol{\mathit{T}} s(x) - \mu(x) \rho(x),
\end{equation}
\noindent where $\epsilon$ is the energy density, $\boldsymbol{\mathit{T}}$ the temperature, $s$ the entropy density, $\mu$ the $x$-dependent chemical potential and $\rho$ the charge density.

\begin{figure}[ht]
\center
\includegraphics[width=0.8 \linewidth]{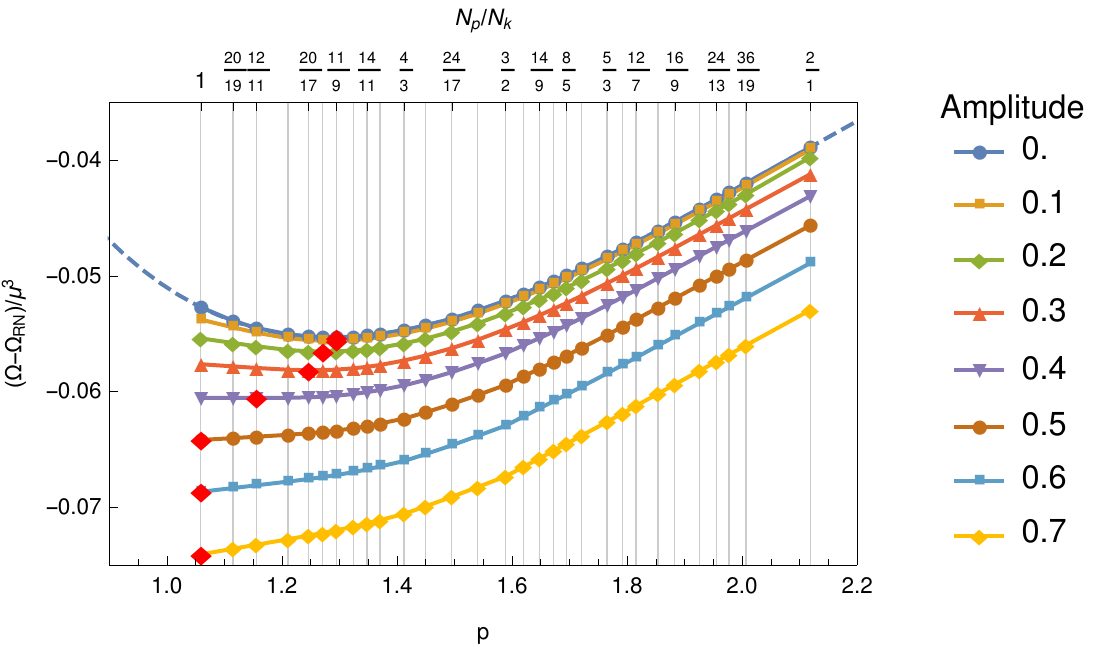}
\caption{\label{fig:wCurves} {\bf The thermodynamic potential of the
    striped solutions on top of the lattice:} Values shown are for lattice momentum
  $k=1.06\mu$ and various amplitude $A$ at
  $T=0.01\mu$. The minima (shown with red diamonds) are shifted from the spontaneous value to the commensurate value as the lattice amplitude is increased. Dashed line shows the energy profile of the spontaneous solution in translationally symmetric background.} 
\end{figure}

There is a peculiar technical difficulty, which arises as soon as one addresses the nonlinear solutions. The problem involves 
two unrelated length scales: the wavelength of the spontaneous structure $\lambda_p = 2\pi/p$ and the wavelength of the background lattice $\lambda_k = 2\pi/k$. 
In order to set up the numerical PDE solver procedure one has to specify only one scale, corresponding to the size of the computational domain with periodic boundary conditions. It is therefore clear that in practice we can only access the values $\lambda_p$ which are rational multiples of $\lambda_k$:
\begin{equation}
\label{eq:rational}
 \lambda_p = \frac{N_k}{N_p} \lambda_k, \qquad N_k, N_p \in \mathbb{N}. 
 \end{equation}
In this case one can choose the computation domain of the size $N_k \lambda_k$ equal to the integer number of lattice periods, which would simultaneously accommodate $N_p$ periods of CDW. This situation is completely analogous to the ``magnetic unit cell'' phenomenon, which arises when one considers a crystal in external magnetic field. The unit cell in this case must simultaneously accommodate integer number of the crystal plaquettes and magnetic fluxes, and can become substantially large \cite{lifshitz2013statistical}. We see here that the accessible range of spontaneous structure wave-vectors $k$ is now discrete and its density is limited by the maximal size of the computational domain, which we can handle in our numerical analysis. In what follows we will use computational domains including up to $N_k =20$ periods of the lattice or up to $N_p = 20$ periods of the spontaneous CDW\footnote{Note once again, that this corresponds to 10 periods in the spontaneous currents}, which allows us to achieve reasonable resolution in our study of the corresponding thermodynamic potentials (see Fig.\ref{fig:wCurves}).

In practice, in order to construct the solution with $N_p$ CDW periods
on top of the $N_k$ lattice cells with period $\lambda_k$ and
amplitude $A$ we first find the spontaneous stripe solution with
specific period $\lambda_p$ from \eqref{eq:rational} on top of the
translationally symmetric background. Then we concatenate $N_p$ copies
of these stripes fitting them in the enlarged calculation domain. At
this point we turn on $N_k$ periods of the background lattice by
slowly changing the boundary condition for the chemical potential,
eventually achieving the desired value of $A$ in \eqref{eq:mu x}. This
adiabatic process preserves the initial number of the CDW periods what
we check numerically at every stage by counting the number of zeros of
the oscillating $A_y$ field at the horizon (see
Figs.\ref{fig:LockProfile}c,d). 

We explore the phase diagram at given temperature by first choosing the period and the amplitude of the explicit ionic lattice. Then we construct a set of nonlinear solutions, corresponding to the spontaneous structures with different wave-vectors $p$ on top of this lattice. We calculate the thermodynamic potential \eqref{eq:free energy} for these solutions and we find the one which is thermodynamically preferred. The sample of the $\Omega(k)$ curves, which we get, is shown on Fig.\,\ref{fig:wCurves}. 

A good check of our calculation is that, at $A=0$ our solutions follow the curve, which one would obtain in the study of the spontaneous striped solutions on the homogeneous RN background \cite{Withers:2013loa,Withers:2013kva,Donos:2013wia}. We have checked that for $c_1=9.9$ our results coincide with Fig.\,2 in \cite{Withers:2013kva}. Even though we have access only to the discrete set of values, they lie on the smooth curves which have a well defined minima. 

One can see that as the amplitude rises, the minimum smoothly shifts
from the incommensurate to commensurate point. Thus by increasing the
amplitude we observe the smooth, at least second order phase
transition. The different commensurate points, which are
thermodynamically stable at different temperatures and charge density
are shown on Fig.\,3a in the main text. 


\subsection{Commensurate state} 
\label{sec:commensurate_state}

\begin{figure}[ht!]
\center
\includegraphics[width=0.88\linewidth]{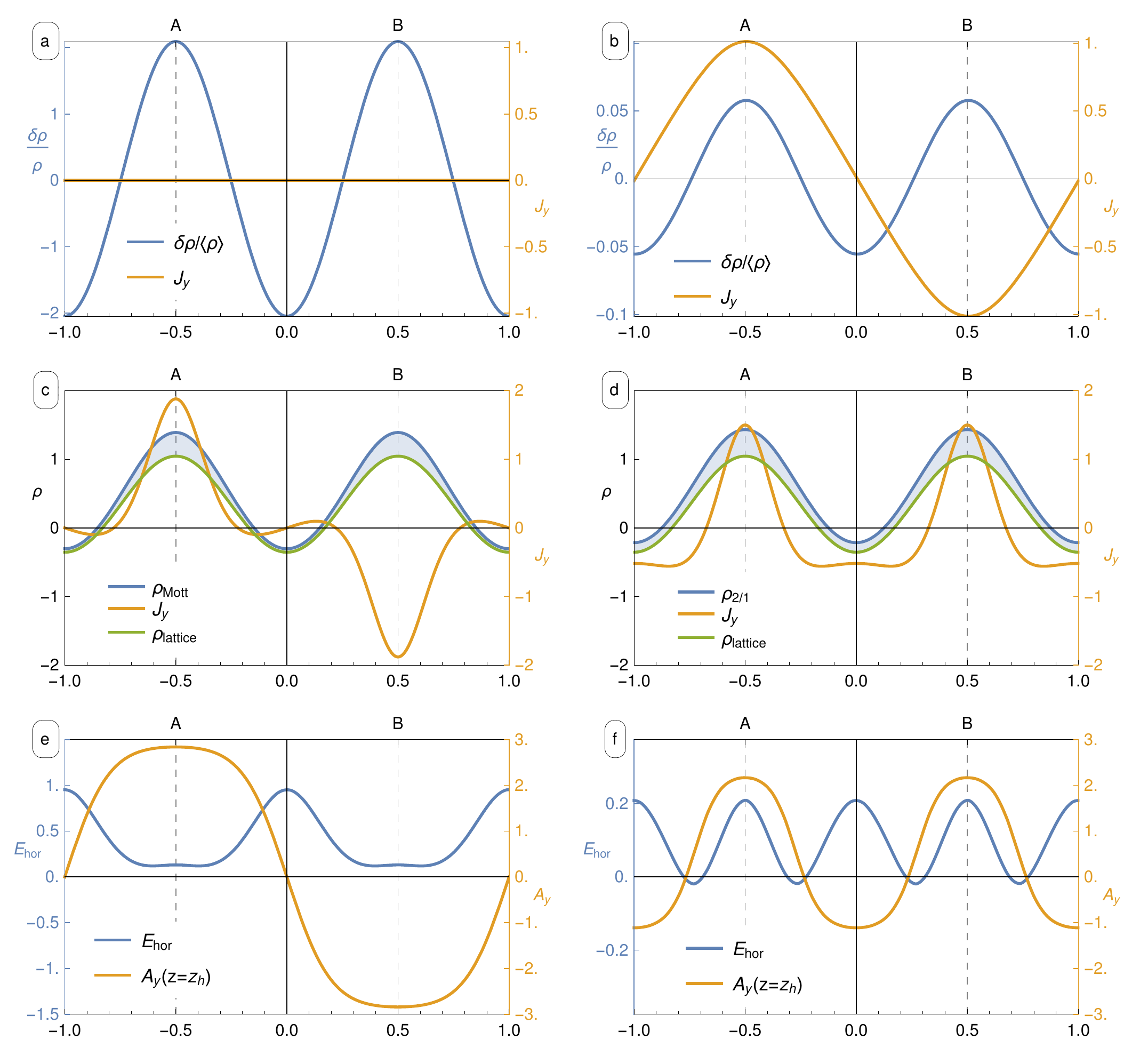}
\caption{\label{fig:LockProfile} \small { \bf Charge and current profiles of
    various states:} (a) The explicit lattice. (b) The spontaneous
  CDW. The modulation of the charge density is strong in the explicit lattice, since it is directly sourced and relatively weak in the pure CDW since it is subleading to the currents order, which drives the instability. \\
(c) The  1/1 commensurate locked state: the Mott insulator, (d) The
2/1 commensurate locked state: the 100\% doped Mott insulator. Note the staggered current pattern in the former case and aligned in the latter. The total current is zero in both cases. The charge density of the lattice in absence of CDW is shown by the green line. The shaded region shows the excess charge due to CDW, which is manifestly positive in both cases.\\ 
(e) The horizon profile in the gravitational theory of the 1/1 state. 
(f) The horizon profile in the gravitational theory of the 2/1 state. Show is the horizon value of $E_{hor}$ of the electric field strength, whose value near UV boundary defines the charge density, and $y$-component of the gauge field $A_y$, related to the current $J_y$ in UV. Near horizon the effect of the lattice is weak and the structure of the locked in CDW (number of its periods) is clearly seen.   
} 
\end{figure}

Let us first focus on two integer value commensurate states: the
leading $1/1$ Mott insulator and the higher $2/1$ commensurate state. 
The features of the $1/1$ state (Fig.\ref{fig:LockProfile}a) are mostly similar to those of the pure spontaneous crystal with an 
important difference: now the periodicity of the overall structure is anchored by the ionic lattice wavelength $\lambda_k$. The staggered currents are seen, but they are now enhanced near the maxima of $\rho(x)$ and suppressed at the minima. This leads to the effective localization of the currents. On the other hand, the spontaneous structure brings an excess of total charge density as compared to the pure lattice (shaded region on Fig.\ref{fig:LockProfile}a, color on Fig.1a-d). This allows us to define a``unit of CDW charge'', $q_{CDW}$ \eqref{equ:qCDW} 
as the integrated excess charge density in the unit cell
\begin{equation}
\label{equ:qCDW}
q_{CDW} \equiv \frac{k}{2 \pi} \int_0^{2 \pi / k} (\rho_{Mott}(x) - \rho_{lattice}(x))
\end{equation}

An important feature is that in holography $q_{CDW}$  assumes continuous range of values depending on the external conditions, while in the hard Mott insulator it would be quantized in units of the electron charge.

Now we address to the higher commensurate state $2/1$ (see Fig.\ref{fig:LockProfile}b). Note that the localized peaks of the $J_y$ current are all aligned. 
The reason is that the $2/1$ commensurate state has twice the number
of CDW periods as compared to the $1/1$ state  (see near horizon
profile on Figs.\ref{fig:LockProfile}c \& d). Every odd positive current peak is thus enhanced by the charge density, but the negative currents are dispersed and do not show well defined peaks. Nonetheless, the total current remains zero. 
Figure \ref{fig:LockProfile}b shows that the total charge density of this ``aligned'' state is larger than that of the staggered one. This is due to the fact that it possess twice the number of spontaneous CDWs per unit cell, each bringing contributions of order $q_{CDW}$ to the total charge density. This feature allows us to denote this state as 100\% doped and define the doping rate as in the main text.

\subsection{Incommensurate state} 
\label{sec:incommensurate_state}
\begin{figure}[ht!]
\center
\includegraphics[width=0.9 \linewidth]{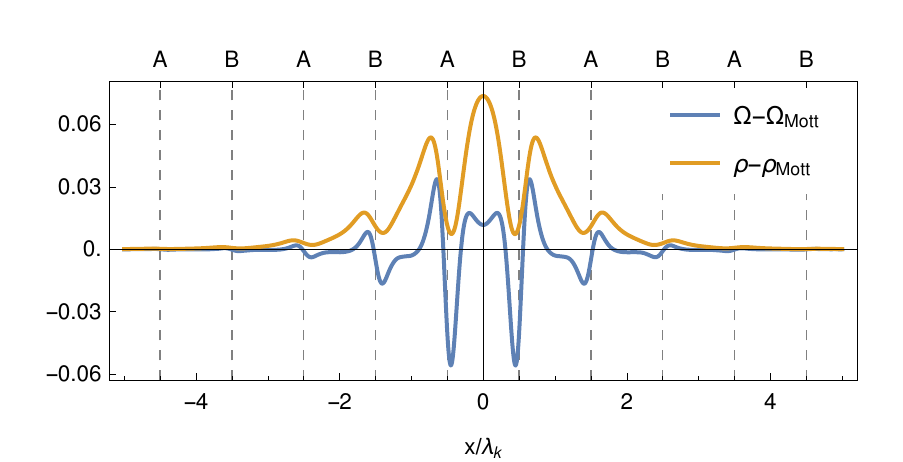}
\caption{\label{fig:DiscProfiles} {\bf Profile of the contribution to
    thermodynamic potential and charge density of a single discommensuration.} This solution has one excess period of the CDW per 19 unit cells, corresponding to the commensurate fraction $p/k = 20/19$ at $A=0.7$, $T=0.01 \mu$. The positive definite total charge of the discommensuration is seen. The size of discommensuration, as obtained from the thermodynamical potential density is about 5 unit cells. All values are measured in units of $\mu$. } 
\end{figure}

As we mentioned earlier, the numerical computation in the incommensurate state is technically more involved, as the numbers of periods in \eqref{eq:rational} can become large.  The isolated discommensuration is found as a solution which is closest to commensurate $\frac{N_p}{N_k} = \frac{1}{1}$ value. Given that
\begin{equation}
\label{discom_counting}
\left| \frac{N_p}{N_k} - \frac{1}{1} \right|= \frac{|N_p - N_k|}{N_k},  
\end{equation}
we will choose $N_k = N_p - 1$ and maximal $N_p$ reachable by our numerics $N_p \leq 20$. The incommensurate solution with 20 CDW's per 19 lattice periods would have exactly one excess CDW period per 19 unit cells as compared to the commensurate state. By inspecting this solution (Fig.\ref{fig:bulk_profiles}) we see, that the solution profile coincides with the commensurate state almost everywhere except from the finite size region in the core, where this excess of one period of CDW is accounted for. We can also study the TD potential and charge density of such solution as compared to the pure commensurate state, Fig.\ref{fig:DiscProfiles}, which shows clearly that this incommensurate solution can be seen as a commensurate state with one localized soliton on top of it. This soliton is a direct analogue of \textit{discommensuration} studied in the context of charge density waves in \cite{mcmillan1976theory,pokrovsky1979ground}.

As it is apparent on Fig.\ref{fig:DiscProfiles} a single discommensuration possess a finite net charge
\begin{equation}
\label{equ:charge_disc}
q_{disc.} = \frac{1}{19 \lambda_k} \int_0^{19 \lambda_k} \rho_{20/19}(x) - \rho_{1/1}(x) dx,
\end{equation}
which is manifestly positive, is of order $q_{CDW}$ \eqref{equ:qCDW} and has a direct analogue in the hard Mott insulator model. Indeed, the discommensurations in the conventional Mott state are associated with one extra or missing electron in the unit cell. In this case the charge of disc. would be exactly $1$. 
The holographic model, on the contrary, allows for continuous variations in $q_{disc.}$, see Fig.\,3c.

One can see that further deviation from the commensurate point $1/1$, according to \eqref{discom_counting}, is achieved by rising the density of discommensurations. At higher commensurate points they can form a super lattice with a, period of several ionic lattice constants $a$ (Fig. 4c). For instance, the discommensuration lattice with $3 a$ period corresponds to the commensurate fraction $p/k=N_p /N_k = 4/3$, which is stable at low temperature in a window of doping levels as shown on the phase diagram Fig.3a.

\subsection{Optical Conductivity} 
\label{sec:conductivity}

In order to further investigate the properties of the above described solutions, we extract their electric conductivity 
as a function of frequency, $\sigma(\omega)$, along the $x$-direction, closely following \cite{Rangamani:2015hka}. 
The optical conductivity in inhomogeneous setups has also been studied in e.g. \cite{Horowitz:2012ky, Horowitz:2012gs, Donos:2014yya}.

Extracting $\sigma(\omega)$ is a considerably involved numerical problem, which requires to first construct a given background to sufficient accuracy and then solve the perturbation equations on top of this solution. We found that our usage of the standard \texttt{MachinePrecision} computations of \textit{Wolfram Mathematica}\cite{Mathematica10} limits the reliability of our AC conductivity results to the region in the vicinity of the critical temperature. For the results within this region we successfully perform a set of numerical consistency checks, which includes vanishing of the constrains and gauge fixing conditions. We have also checked that our results satisfy the sum rule on the integrated spectral weight: $\lim_{\omega \rightarrow 0} S(\omega) \rightarrow 0$, where
\begin{equation}
S(\omega/\mu) \equiv \int_0^{\omega/\mu}(Re[\sigma(\omega')]-1)d\omega'.
\end{equation}
If one is interested only in the DC conductivity, $\sigma_{DC} = \sigma(\omega = 0)$, it is possible to largely simplify 
the calculation since we can obtain a formula for this observable solely in terms of the horizon data 
\cite{Donos:2014yya,Donos:2015gia, Banks:2015wha, Donos:2015bxe, Donos:2017mhp}.
The relevant formula for our system was obtained in \cite{Donos:2017mhp}, which we have rederived in full agreement. 
Additionally, for all the cases we have studied, the limit $\omega \to 0^+$ of $\sigma(\omega)$ agrees with the 
computation of $\sigma_{DC}$ in terms of the horizon data, which serves as another test of our numerics. 

\subsection{Numerical techniques and Precision control} 
\label{sec:numerical_techniques}

In the present study we rely heavily on numerical analysis of the holographic nonlinear solutions. Moreover, in order to study the phase diagram and cover the parameter space we have to obtain several thousands of solutions, with some of them requiring quite large calculation grids in the spatial direction. This situation places very strict requirements on the numerical techniques which we use, the precision and the accuracy of the results. 

We have chosen a single patch pseudospectral scheme in the holographic direction
and used \textit{Wolfram Mathematica}\cite{Mathematica10} to implement the numerical algorithm. 
The main limitation which we encounter is the necessity to work with \texttt{MachinePrecision} numbers in the compiled function, which eventually limits the precision of our results. 
Element-wise operations can be efficiently compiled with \texttt{Compile}, which brings up a spectacular acceleration.
We use precompiled linear algebra solvers and sparse matrices, which delivers decent speed of calculations.

It should be noted that direct inversion (Newton-Raphson method) for the case of pseudospectral discretization is 
extremely demanding for large grids, so we have decided to use a relaxation scheme instead.
We employ the differential operator evaluated in the low order finite difference derivative scheme as a preconditioner. The result is a nonlinear Richardson relaxation with Orszag preconditioning (See Sec. 15.14 and eq.(15.115) in \cite{boyd2001chebyshev}).
At the end of the day, we managed to optimize the calculation scheme to the extent that it takes about half an hour to obtain the precise solution on our largest grid of size $\sim 330_x \times 80_z$ (pseudospectral) using a single core of a laptop CPU (Intel Core i7-5600U @ 2.60GHz ) and about 3 Gb of RAM.

As one can see from our results, the difference between the free energies of the solution with spontaneous structure and the one without 
is just of order of a few per cents of the free energies themselves. This means that in order to reliably study this difference, we need to evaluate the free energies with accuracy of at least $10^{-4}$. 
We observe that for a single patch Chebyshev grid the maximum $N_y$ resolution is limited by the rounding errors at $N_y=80$. The accuracy of the thermodynamical potential for a grid of this size is about $10^{-7}$. We used this value as a numerical error estimate throughout the present study and it has proven to be sufficient to obtain our main results.
One should keep in mind that in the numerical procedure we solve the modified DeTurk equations. Thus, it must be checked that the Einstein equations are satisfied, which we do by two independent measures:  the maximal value of the trace of the Einstein equations and the maximum value of the norm of a DeTurck vector. For 
temperatures $T> 0.01$, these values are both of order $10^{-7}$, which is quite satisfactory \cite{Donos:2014yya}.

\bibliography{inhom_stripes_lattice}

\begin{thebibliography}{10}
\expandafter\ifx\csname url\endcsname\relax
  \def\url#1{\texttt{#1}}\fi
\expandafter\ifx\csname urlprefix\endcsname\relax\def\urlprefix{URL }\fi
\providecommand{\bibinfo}[2]{#2}
\providecommand{\eprint}[2][]{\url{#2}}

\bibitem{zaanen1985band}
\bibinfo{author}{Zaanen, J.}, \bibinfo{author}{Sawatzky, G.} \&
  \bibinfo{author}{Allen, J.}
\newblock \bibinfo{title}{Band gaps and electronic structure of
  transition-metal compounds}.
\newblock \emph{\bibinfo{journal}{Physical Review Letters}}
  \textbf{\bibinfo{volume}{55}}, \bibinfo{pages}{418} (\bibinfo{year}{1985}).

\bibitem{fradkin2013field}
\bibinfo{author}{Fradkin, E.}
\newblock \emph{\bibinfo{title}{Field theories of condensed matter physics}}
  (\bibinfo{publisher}{Cambridge University Press}, \bibinfo{year}{2013}).

\bibitem{wen2004quantum}
\bibinfo{author}{Wen, X.-G.}
\newblock \emph{\bibinfo{title}{Quantum field theory of many-body systems: from
  the origin of sound to an origin of light and electrons}}
  (\bibinfo{publisher}{Oxford University Press}, \bibinfo{year}{2004}).

\bibitem{zaanen2008pacifying}
\bibinfo{author}{Zaanen, J.}, \bibinfo{author}{Krueger, F.},
  \bibinfo{author}{She, J.}, \bibinfo{author}{Sadri, D.} \&
  \bibinfo{author}{Mukhin, S.}
\newblock \bibinfo{title}{Pacifying the fermi-liquid: battling the devious
  fermion signs}.
\newblock \emph{\bibinfo{journal}{Iranian Journal of Physics Research}}
  \textbf{\bibinfo{volume}{8}}, \bibinfo{pages}{111--111}
  (\bibinfo{year}{2008}).

\bibitem{Zaanen:2015oix}
\bibinfo{author}{Zaanen, J.}, \bibinfo{author}{Sun, Y.-W.},
  \bibinfo{author}{Liu, Y.} \& \bibinfo{author}{Schalm, K.}
\newblock \emph{\bibinfo{title}{{Holographic Duality in Condensed Matter
  Physics}}} (\bibinfo{publisher}{Cambridge Univ. Press},
  \bibinfo{year}{2015}).
\newblock
  \urlprefix\url{http://www.cambridge.org/mw/academic/subjects/physics/condensed-matter-physics-nanoscience-and-mesoscopic-physics/holographic-duality-condensed-matter-physics?format=HB}.

\bibitem{keimer2015quantum}
\bibinfo{author}{Keimer, B.}, \bibinfo{author}{Kivelson, S.},
  \bibinfo{author}{Norman, M.}, \bibinfo{author}{Uchida, S.} \&
  \bibinfo{author}{Zaanen, J.}
\newblock \bibinfo{title}{From quantum matter to high-temperature
  superconductivity in copper oxides}.
\newblock \emph{\bibinfo{journal}{Nature}} \textbf{\bibinfo{volume}{518}},
  \bibinfo{pages}{179} (\bibinfo{year}{2015}).

\bibitem{fradkin2015colloquium}
\bibinfo{author}{Fradkin, E.}, \bibinfo{author}{Kivelson, S.~A.} \&
  \bibinfo{author}{Tranquada, J.~M.}
\newblock \bibinfo{title}{Colloquium: Theory of intertwined orders in high
  temperature superconductors}.
\newblock \emph{\bibinfo{journal}{Reviews of Modern Physics}}
  \textbf{\bibinfo{volume}{87}}, \bibinfo{pages}{457} (\bibinfo{year}{2015}).

\bibitem{mesaros2016commensurate}
\bibinfo{author}{Mesaros, A.} \emph{et~al.}
\newblock \bibinfo{title}{Commensurate 4a0-period charge density modulations
  throughout the bi2sr2cacu2o8+ x pseudogap regime}.
\newblock \emph{\bibinfo{journal}{Proceedings of the National Academy of
  Sciences}} \textbf{\bibinfo{volume}{113}}, \bibinfo{pages}{12661--12666}
  (\bibinfo{year}{2016}).

\bibitem{ammon2015gauge}
\bibinfo{author}{Ammon, M.} \& \bibinfo{author}{Erdmenger, J.}
\newblock \emph{\bibinfo{title}{Gauge/gravity duality: foundations and
  applications}} (\bibinfo{publisher}{Cambridge University Press},
  \bibinfo{year}{2015}).

\bibitem{faulkner2010strange}
\bibinfo{author}{Faulkner, T.}, \bibinfo{author}{Iqbal, N.},
  \bibinfo{author}{Liu, H.}, \bibinfo{author}{McGreevy, J.} \&
  \bibinfo{author}{Vegh, D.}
\newblock \bibinfo{title}{Strange metal transport realized by gauge/gravity
  duality}.
\newblock \emph{\bibinfo{journal}{Science}} \textbf{\bibinfo{volume}{329}},
  \bibinfo{pages}{1043--1047} (\bibinfo{year}{2010}).

\bibitem{iqbal2011lectures}
\bibinfo{author}{Iqbal, N.}, \bibinfo{author}{Liu, H.} \&
  \bibinfo{author}{Mezei, M.}
\newblock \bibinfo{title}{Lectures on holographic non-fermi liquids and quantum
  phase transitions}  (\bibinfo{year}{2011}).
\newblock \eprint{1110.3814}.

\bibitem{Policastro:2001yc}
\bibinfo{author}{Policastro, G.}, \bibinfo{author}{Son, D.~T.} \&
  \bibinfo{author}{Starinets, A.~O.}
\newblock \bibinfo{title}{{The Shear viscosity of strongly coupled N=4
  supersymmetric Yang-Mills plasma}}.
\newblock \emph{\bibinfo{journal}{Phys. Rev. Lett.}}
  \textbf{\bibinfo{volume}{87}}, \bibinfo{pages}{081601}
  (\bibinfo{year}{2001}).
\newblock \eprint{hep-th/0104066}.

\bibitem{hartnoll2016holographic}
\bibinfo{author}{Hartnoll, S.~A.}, \bibinfo{author}{Lucas, A.} \&
  \bibinfo{author}{Sachdev, S.}
\newblock \bibinfo{title}{Holographic quantum matter}  (\bibinfo{year}{2016}).
\newblock \eprint{1612.07324}.

\bibitem{zaanen1990systematics}
\bibinfo{author}{Zaanen, J.} \& \bibinfo{author}{Sawatzky, G.}
\newblock \bibinfo{title}{Systematics in band gaps and optical spectra of 3d
  transition metal compounds}.
\newblock \emph{\bibinfo{journal}{Journal of solid state chemistry}}
  \textbf{\bibinfo{volume}{88}}, \bibinfo{pages}{8--27} (\bibinfo{year}{1990}).

\bibitem{rozenberg1995optical}
\bibinfo{author}{Rozenberg, M.} \emph{et~al.}
\newblock \bibinfo{title}{Optical conductivity in mott-hubbard systems}.
\newblock \emph{\bibinfo{journal}{Physical review letters}}
  \textbf{\bibinfo{volume}{75}}, \bibinfo{pages}{105} (\bibinfo{year}{1995}).

\bibitem{anderson1950antiferromagnetism}
\bibinfo{author}{Anderson, P.}
\newblock \bibinfo{title}{Antiferromagnetism. theory of superexchange
  interaction}.
\newblock \emph{\bibinfo{journal}{Physical Review}}
  \textbf{\bibinfo{volume}{79}}, \bibinfo{pages}{350} (\bibinfo{year}{1950}).

\bibitem{zaanen1987electronic}
\bibinfo{author}{Zaanen, J.} \& \bibinfo{author}{Sawatzky, G.}
\newblock \bibinfo{title}{The electronic structure and superexchange
  interactions in transition-metal compounds}.
\newblock \emph{\bibinfo{journal}{Canadian journal of physics}}
  \textbf{\bibinfo{volume}{65}}, \bibinfo{pages}{1262--1271}
  (\bibinfo{year}{1987}).

\bibitem{zaanen1989charged}
\bibinfo{author}{Zaanen, J.} \& \bibinfo{author}{Gunnarsson, O.}
\newblock \bibinfo{title}{Charged magnetic domain lines and the magnetism of
  high-t c oxides}.
\newblock \emph{\bibinfo{journal}{Physical Review B}}
  \textbf{\bibinfo{volume}{40}}, \bibinfo{pages}{7391} (\bibinfo{year}{1989}).

\bibitem{tranquada1995evidence}
\bibinfo{author}{Tranquada, J.}, \bibinfo{author}{Sternlieb, B.},
  \bibinfo{author}{Axe, J.}, \bibinfo{author}{Nakamura, Y.} \&
  \bibinfo{author}{Uchida, S.}
\newblock \bibinfo{title}{Evidence for stripe correlations of spins and holes
  in copper oxide superconductors}.
\newblock \emph{\bibinfo{journal}{Nature}} \textbf{\bibinfo{volume}{375}},
  \bibinfo{pages}{561} (\bibinfo{year}{1995}).

\bibitem{vojta2009lattice}
\bibinfo{author}{Vojta, M.}
\newblock \bibinfo{title}{Lattice symmetry breaking in cuprate superconductors:
  stripes, nematics, and superconductivity}.
\newblock \emph{\bibinfo{journal}{Advances in Physics}}
  \textbf{\bibinfo{volume}{58}}, \bibinfo{pages}{699--820}
  (\bibinfo{year}{2009}).

\bibitem{zheng2016stripe}
\bibinfo{author}{Zheng, B.-X.} \emph{et~al.}
\newblock \bibinfo{title}{Stripe order in the underdoped region of the
  two-dimensional hubbard model}  (\bibinfo{year}{2016}).
\newblock \eprint{1701.00054}.

\bibitem{huang2016numerical}
\bibinfo{author}{Huang, E.~W.} \emph{et~al.}
\newblock \bibinfo{title}{Numerical evidence of fluctuating stripes in the
  normal state of high-tc cuprate superconductors}  (\bibinfo{year}{2016}).
\newblock \eprint{1612.05211}.

\bibitem{Donos:2013gda}
\bibinfo{author}{Donos, A.} \& \bibinfo{author}{Gauntlett, J.~P.}
\newblock \bibinfo{title}{{Holographic charge density waves}}.
\newblock \emph{\bibinfo{journal}{Phys. Rev.}} \textbf{\bibinfo{volume}{D87}},
  \bibinfo{pages}{126008} (\bibinfo{year}{2013}).
\newblock \eprint{1303.4398}.

\bibitem{fauque2006magnetic}
\bibinfo{author}{Fauqu{\'e}, B.} \emph{et~al.}
\newblock \bibinfo{title}{Magnetic order in the pseudogap phase of high-t c
  superconductors}.
\newblock \emph{\bibinfo{journal}{Physical Review Letters}}
  \textbf{\bibinfo{volume}{96}}, \bibinfo{pages}{197001}
  (\bibinfo{year}{2006}).

\bibitem{li2008unusual}
\bibinfo{author}{Li, Y.} \emph{et~al.}
\newblock \bibinfo{title}{Unusual magnetic order in the pseudogap region of the
  superconductor hgba\^{} sub 2\^{} cuo\^{} sub 4+[delta]\^{}}.
\newblock \emph{\bibinfo{journal}{Nature}} \textbf{\bibinfo{volume}{455}},
  \bibinfo{pages}{372} (\bibinfo{year}{2008}).

\bibitem{li2010hidden}
\bibinfo{author}{{Li Yuan}} \emph{et~al.}
\newblock \bibinfo{title}{{Hidden magnetic excitation in the pseudogap phase of
  a high-Tc superconductor}}.
\newblock \emph{\bibinfo{journal}{Nature}} \textbf{\bibinfo{volume}{468}},
  \bibinfo{pages}{283--285} (\bibinfo{year}{2010}).
\newblock
  \urlprefix\url{http://www.nature.com/nature/journal/v468/n7321/abs/nature09477.html\#supplementary-information}.

\bibitem{Zhao2017global}
\bibinfo{author}{{Zhao L.}} \emph{et~al.}
\newblock \bibinfo{title}{{A global inversion-symmetry-broken phase inside the
  pseudogap region of YBa2Cu3Oy}}.
\newblock \emph{\bibinfo{journal}{Nature Physics}}
  \textbf{\bibinfo{volume}{13}}, \bibinfo{pages}{250--254}
  (\bibinfo{year}{2017}).
\newblock
  \urlprefix\url{http://www.nature.com/nphys/journal/v13/n3/abs/nphys3962.html\#supplementary-information}.

\bibitem{li2007two}
\bibinfo{author}{Li, Q.}, \bibinfo{author}{H{\"u}cker, M.},
  \bibinfo{author}{Gu, G.}, \bibinfo{author}{Tsvelik, A.} \&
  \bibinfo{author}{Tranquada, J.}
\newblock \bibinfo{title}{Two-dimensional superconducting fluctuations in
  stripe-ordered la 1.875 ba 0.125 cuo 4}.
\newblock \emph{\bibinfo{journal}{Physical review letters}}
  \textbf{\bibinfo{volume}{99}}, \bibinfo{pages}{067001}
  (\bibinfo{year}{2007}).

\bibitem{rajasekaran2017probing}
\bibinfo{author}{Rajasekaran, S.} \emph{et~al.}
\newblock \bibinfo{title}{Probing optically silent superfluid stripes in
  cuprates}  (\bibinfo{year}{2017}).
\newblock \eprint{1705.06112}.

\bibitem{hamidian2016detection}
\bibinfo{author}{{Hamidian M. H.}} \emph{et~al.}
\newblock \bibinfo{title}{{Detection of a Cooper-pair density wave in
  Bi2Sr2CaCu2O8+x}}.
\newblock \emph{\bibinfo{journal}{Nature}} \textbf{\bibinfo{volume}{532}},
  \bibinfo{pages}{343--347} (\bibinfo{year}{2016}).

\bibitem{Ooguri:2010kt}
\bibinfo{author}{Ooguri, H.} \& \bibinfo{author}{Park, C.-S.}
\newblock \bibinfo{title}{{Holographic End-Point of Spatially Modulated Phase
  Transition}}.
\newblock \emph{\bibinfo{journal}{Phys. Rev.}} \textbf{\bibinfo{volume}{D82}},
  \bibinfo{pages}{126001} (\bibinfo{year}{2010}).
\newblock \eprint{1007.3737}.

\bibitem{Donos:2011bh}
\bibinfo{author}{Donos, A.} \& \bibinfo{author}{Gauntlett, J.~P.}
\newblock \bibinfo{title}{{Holographic striped phases}}.
\newblock \emph{\bibinfo{journal}{JHEP}} \textbf{\bibinfo{volume}{08}},
  \bibinfo{pages}{140} (\bibinfo{year}{2011}).
\newblock \eprint{1106.2004}.

\bibitem{Cai:2017qdz}
\bibinfo{author}{Cai, R.-G.}, \bibinfo{author}{Li, L.}, \bibinfo{author}{Wang,
  Y.-Q.} \& \bibinfo{author}{Zaanen, J.}
\newblock \bibinfo{title}{{Intertwined order and holography: the case of the
  pair density wave}}.
\newblock \emph{\bibinfo{journal}{arXiv : 1706.01470}}  (\bibinfo{year}{2017}).
\newblock \eprint{1706.01470}.

\bibitem{Withers:2014sja}
\bibinfo{author}{Withers, B.}
\newblock \bibinfo{title}{{Holographic Checkerboards}}.
\newblock \emph{\bibinfo{journal}{JHEP}} \textbf{\bibinfo{volume}{09}},
  \bibinfo{pages}{102} (\bibinfo{year}{2014}).
\newblock \eprint{1407.1085}.

\bibitem{Flauger:2010tv}
\bibinfo{author}{Flauger, R.}, \bibinfo{author}{Pajer, E.} \&
  \bibinfo{author}{Papanikolaou, S.}
\newblock \bibinfo{title}{{A Striped Holographic Superconductor}}.
\newblock \emph{\bibinfo{journal}{Phys. Rev.}} \textbf{\bibinfo{volume}{D83}},
  \bibinfo{pages}{064009} (\bibinfo{year}{2011}).
\newblock \eprint{1010.1775}.

\bibitem{Liu:2012tr}
\bibinfo{author}{Liu, Y.}, \bibinfo{author}{Schalm, K.}, \bibinfo{author}{Sun,
  Y.-W.} \& \bibinfo{author}{Zaanen, J.}
\newblock \bibinfo{title}{{Lattice Potentials and Fermions in Holographic non
  Fermi-Liquids: Hybridizing Local Quantum Criticality}}.
\newblock \emph{\bibinfo{journal}{JHEP}} \textbf{\bibinfo{volume}{10}},
  \bibinfo{pages}{036} (\bibinfo{year}{2012}).
\newblock \eprint{1205.5227}.

\bibitem{Horowitz:2012ky}
\bibinfo{author}{Horowitz, G.~T.}, \bibinfo{author}{Santos, J.~E.} \&
  \bibinfo{author}{Tong, D.}
\newblock \bibinfo{title}{{Optical Conductivity with Holographic Lattices}}.
\newblock \emph{\bibinfo{journal}{JHEP}} \textbf{\bibinfo{volume}{07}},
  \bibinfo{pages}{168} (\bibinfo{year}{2012}).
\newblock \eprint{1204.0519}.

\bibitem{Horowitz:2012gs}
\bibinfo{author}{Horowitz, G.~T.}, \bibinfo{author}{Santos, J.~E.} \&
  \bibinfo{author}{Tong, D.}
\newblock \bibinfo{title}{{Further Evidence for Lattice-Induced Scaling}}.
\newblock \emph{\bibinfo{journal}{JHEP}} \textbf{\bibinfo{volume}{11}},
  \bibinfo{pages}{102} (\bibinfo{year}{2012}).
\newblock \eprint{1209.1098}.

\bibitem{Donos:2014yya}
\bibinfo{author}{Donos, A.} \& \bibinfo{author}{Gauntlett, J.~P.}
\newblock \bibinfo{title}{{The thermoelectric properties of inhomogeneous
  holographic lattices}}.
\newblock \emph{\bibinfo{journal}{JHEP}} \textbf{\bibinfo{volume}{01}},
  \bibinfo{pages}{035} (\bibinfo{year}{2015}).
\newblock \eprint{1409.6875}.

\bibitem{Rangamani:2015hka}
\bibinfo{author}{Rangamani, M.}, \bibinfo{author}{Rozali, M.} \&
  \bibinfo{author}{Smyth, D.}
\newblock \bibinfo{title}{{Spatial Modulation and Conductivities in Effective
  Holographic Theories}}.
\newblock \emph{\bibinfo{journal}{JHEP}} \textbf{\bibinfo{volume}{07}},
  \bibinfo{pages}{024} (\bibinfo{year}{2015}).
\newblock \eprint{1505.05171}.

\bibitem{pokrovsky1979ground}
\bibinfo{author}{Pokrovsky, V.} \& \bibinfo{author}{Talapov, A.}
\newblock \bibinfo{title}{Ground state, spectrum, and phase diagram of
  two-dimensional incommensurate crystals}.
\newblock \emph{\bibinfo{journal}{Physical Review Letters}}
  \textbf{\bibinfo{volume}{42}}, \bibinfo{pages}{65} (\bibinfo{year}{1979}).

\bibitem{bak1982commensurate}
\bibinfo{author}{Bak, P.}
\newblock \bibinfo{title}{Commensurate phases, incommensurate phases and the
  devil's staircase}.
\newblock \emph{\bibinfo{journal}{Reports on Progress in Physics}}
  \textbf{\bibinfo{volume}{45}}, \bibinfo{pages}{587} (\bibinfo{year}{1982}).

\bibitem{Andrade:2017leb}
\bibinfo{author}{Andrade, T.} \& \bibinfo{author}{Krikun, A.}
\newblock \bibinfo{title}{{Commensurate lock-in in holographic non-homogeneous
  lattices}}.
\newblock \emph{\bibinfo{journal}{JHEP}} \textbf{\bibinfo{volume}{03}},
  \bibinfo{pages}{168} (\bibinfo{year}{2017}).
\newblock \eprint{1701.04625}.

\bibitem{FKbook}
\bibinfo{author}{Braun, O.} \& \bibinfo{author}{Kivshar, Y.}
\newblock \emph{\bibinfo{title}{The Frenkel-Kontorova Model: Concepts, Methods
  and Applications}} (\bibinfo{publisher}{Springer-Verlag Berlin Heidelberg},
  \bibinfo{year}{2004}).

\bibitem{comin2016resonant}
\bibinfo{author}{Comin, R.} \& \bibinfo{author}{Damascelli, A.}
\newblock \bibinfo{title}{Resonant x-ray scattering studies of charge order in
  cuprates}.
\newblock \emph{\bibinfo{journal}{Annual Review of Condensed Matter Physics}}
  \textbf{\bibinfo{volume}{7}}, \bibinfo{pages}{369--405}
  (\bibinfo{year}{2016}).

\bibitem{boebinger1996insulator}
\bibinfo{author}{Boebinger, G.} \emph{et~al.}
\newblock \bibinfo{title}{Insulator-to-metal crossover in the normal state of
  la 2- x sr x cuo 4 near optimum doping}.
\newblock \emph{\bibinfo{journal}{Physical Review Letters}}
  \textbf{\bibinfo{volume}{77}}, \bibinfo{pages}{5417} (\bibinfo{year}{1996}).

\bibitem{laliberte2016origin}
\bibinfo{author}{Laliberte, F.} \emph{et~al.}
\newblock \bibinfo{title}{Origin of the metal-to-insulator crossover in cuprate
  superconductors}  (\bibinfo{year}{2016}).
\newblock \eprint{1606.04491}.

\bibitem{Grozdanov:2015qia}
\bibinfo{author}{Grozdanov, S.}, \bibinfo{author}{Lucas, A.},
  \bibinfo{author}{Sachdev, S.} \& \bibinfo{author}{Schalm, K.}
\newblock \bibinfo{title}{{Absence of disorder-driven metal-insulator
  transitions in simple holographic models}}.
\newblock \emph{\bibinfo{journal}{Phys. Rev. Lett.}}
  \textbf{\bibinfo{volume}{115}}, \bibinfo{pages}{221601}
  (\bibinfo{year}{2015}).
\newblock \eprint{1507.00003}.

\bibitem{Donos:2014oha}
\bibinfo{author}{Donos, A.}, \bibinfo{author}{Goutéraux, B.} \&
  \bibinfo{author}{Kiritsis, E.}
\newblock \bibinfo{title}{{Holographic Metals and Insulators with Helical
  Symmetry}}.
\newblock \emph{\bibinfo{journal}{JHEP}} \textbf{\bibinfo{volume}{09}},
  \bibinfo{pages}{038} (\bibinfo{year}{2014}).
\newblock \eprint{1406.6351}.

\bibitem{Withers:2013kva}
\bibinfo{author}{Withers, B.}
\newblock \bibinfo{title}{{The moduli space of striped black branes}}.
\newblock \emph{\bibinfo{journal}{arXiv: 1304.2011}}  (\bibinfo{year}{2013}).
\newblock \eprint{1304.2011}.

\bibitem{Withers:2013loa}
\bibinfo{author}{Withers, B.}
\newblock \bibinfo{title}{{Black branes dual to striped phases}}.
\newblock \emph{\bibinfo{journal}{Class. Quant. Grav.}}
  \textbf{\bibinfo{volume}{30}}, \bibinfo{pages}{155025}
  (\bibinfo{year}{2013}).
\newblock \eprint{1304.0129}.

\bibitem{deHaro:2000vlm}
\bibinfo{author}{de~Haro, S.}, \bibinfo{author}{Solodukhin, S.~N.} \&
  \bibinfo{author}{Skenderis, K.}
\newblock \bibinfo{title}{{Holographic reconstruction of space-time and
  renormalization in the AdS / CFT correspondence}}.
\newblock \emph{\bibinfo{journal}{Commun. Math. Phys.}}
  \textbf{\bibinfo{volume}{217}}, \bibinfo{pages}{595--622}
  (\bibinfo{year}{2001}).
\newblock \eprint{hep-th/0002230}.

\bibitem{Maldacena:1997re}
\bibinfo{author}{Maldacena, J.~M.}
\newblock \bibinfo{title}{{The Large N limit of superconformal field theories
  and supergravity}}.
\newblock \emph{\bibinfo{journal}{Int. J. Theor. Phys.}}
  \textbf{\bibinfo{volume}{38}}, \bibinfo{pages}{1113--1133}
  (\bibinfo{year}{1999}).
\newblock \bibinfo{note}{[Adv. Theor. Math. Phys.2,231(1998)]},
  \eprint{hep-th/9711200}.

\bibitem{Witten:1998qj}
\bibinfo{author}{Witten, E.}
\newblock \bibinfo{title}{{Anti-de Sitter space and holography}}.
\newblock \emph{\bibinfo{journal}{Adv. Theor. Math. Phys.}}
  \textbf{\bibinfo{volume}{2}}, \bibinfo{pages}{253--291}
  (\bibinfo{year}{1998}).
\newblock \eprint{hep-th/9802150}.

\bibitem{Gubser:1998bc}
\bibinfo{author}{Gubser, S.~S.}, \bibinfo{author}{Klebanov, I.~R.} \&
  \bibinfo{author}{Polyakov, A.~M.}
\newblock \bibinfo{title}{{Gauge theory correlators from noncritical string
  theory}}.
\newblock \emph{\bibinfo{journal}{Phys. Lett.}}
  \textbf{\bibinfo{volume}{B428}}, \bibinfo{pages}{105--114}
  (\bibinfo{year}{1998}).
\newblock \eprint{hep-th/9802109}.

\bibitem{Donos:2013wia}
\bibinfo{author}{Donos, A.}
\newblock \bibinfo{title}{{Striped phases from holography}}.
\newblock \emph{\bibinfo{journal}{JHEP}} \textbf{\bibinfo{volume}{05}},
  \bibinfo{pages}{059} (\bibinfo{year}{2013}).
\newblock \eprint{1303.7211}.

\bibitem{Gauntlett:2009bh}
\bibinfo{author}{Gauntlett, J.~P.}, \bibinfo{author}{Sonner, J.} \&
  \bibinfo{author}{Wiseman, T.}
\newblock \bibinfo{title}{{Quantum Criticality and Holographic Superconductors
  in M-theory}}.
\newblock \emph{\bibinfo{journal}{JHEP}} \textbf{\bibinfo{volume}{02}},
  \bibinfo{pages}{060} (\bibinfo{year}{2010}).
\newblock \eprint{0912.0512}.

\bibitem{Rozali:2012es}
\bibinfo{author}{Rozali, M.}, \bibinfo{author}{Smyth, D.},
  \bibinfo{author}{Sorkin, E.} \& \bibinfo{author}{Stang, J.~B.}
\newblock \bibinfo{title}{{Holographic Stripes}}.
\newblock \emph{\bibinfo{journal}{Phys. Rev. Lett.}}
  \textbf{\bibinfo{volume}{110}}, \bibinfo{pages}{201603}
  (\bibinfo{year}{2013}).
\newblock \eprint{1211.5600}.

\bibitem{Nakamura:2009tf}
\bibinfo{author}{Nakamura, S.}, \bibinfo{author}{Ooguri, H.} \&
  \bibinfo{author}{Park, C.-S.}
\newblock \bibinfo{title}{{Gravity Dual of Spatially Modulated Phase}}.
\newblock \emph{\bibinfo{journal}{Phys. Rev.}} \textbf{\bibinfo{volume}{D81}},
  \bibinfo{pages}{044018} (\bibinfo{year}{2010}).
\newblock \eprint{0911.0679}.

\bibitem{lifshitz2013statistical}
\bibinfo{author}{Lifshitz, E.~M.} \& \bibinfo{author}{Pitaevskii, L.~P.}
\newblock \emph{\bibinfo{title}{Statistical physics: theory of the condensed
  state}}, vol.~\bibinfo{volume}{9} (\bibinfo{publisher}{Elsevier},
  \bibinfo{year}{2013}).

\bibitem{mcmillan1976theory}
\bibinfo{author}{McMillan, W.}
\newblock \bibinfo{title}{Theory of discommensurations and the
  commensurate-incommensurate charge-density-wave phase transition}.
\newblock \emph{\bibinfo{journal}{Physical Review B}}
  \textbf{\bibinfo{volume}{14}}, \bibinfo{pages}{1496} (\bibinfo{year}{1976}).

\bibitem{Mathematica10}
\bibinfo{author}{{Wolfram Research, Inc.}}
\newblock \emph{\bibinfo{title}{Mathematica, Version 10.2}}
  (\bibinfo{address}{Champaign, Illinois}, \bibinfo{year}{2015}).

\bibitem{Donos:2015gia}
\bibinfo{author}{Donos, A.} \& \bibinfo{author}{Gauntlett, J.~P.}
\newblock \bibinfo{title}{{Navier-Stokes Equations on Black Hole Horizons and
  DC Thermoelectric Conductivity}}.
\newblock \emph{\bibinfo{journal}{Phys. Rev.}} \textbf{\bibinfo{volume}{D92}},
  \bibinfo{pages}{121901} (\bibinfo{year}{2015}).
\newblock \eprint{1506.01360}.

\bibitem{Banks:2015wha}
\bibinfo{author}{Banks, E.}, \bibinfo{author}{Donos, A.} \&
  \bibinfo{author}{Gauntlett, J.~P.}
\newblock \bibinfo{title}{{Thermoelectric DC conductivities and Stokes flows on
  black hole horizons}}.
\newblock \emph{\bibinfo{journal}{JHEP}} \textbf{\bibinfo{volume}{10}},
  \bibinfo{pages}{103} (\bibinfo{year}{2015}).
\newblock \eprint{1507.00234}.

\bibitem{Donos:2015bxe}
\bibinfo{author}{Donos, A.}, \bibinfo{author}{Gauntlett, J.~P.},
  \bibinfo{author}{Griffin, T.} \& \bibinfo{author}{Melgar, L.}
\newblock \bibinfo{title}{{DC Conductivity of Magnetised Holographic Matter}}.
\newblock \emph{\bibinfo{journal}{JHEP}} \textbf{\bibinfo{volume}{01}},
  \bibinfo{pages}{113} (\bibinfo{year}{2016}).
\newblock \eprint{1511.00713}.

\bibitem{Donos:2017mhp}
\bibinfo{author}{Donos, A.}, \bibinfo{author}{Gauntlett, J.~P.},
  \bibinfo{author}{Griffin, T.}, \bibinfo{author}{Lohitsiri, N.} \&
  \bibinfo{author}{Melgar, L.}
\newblock \bibinfo{title}{{Holographic DC conductivity and Onsager relations}}.
\newblock \emph{\bibinfo{journal}{JHEP}} \textbf{\bibinfo{volume}{07}},
  \bibinfo{pages}{006} (\bibinfo{year}{2017}).
\newblock \eprint{1704.05141}.

\bibitem{boyd2001chebyshev}
\bibinfo{author}{Boyd, J.~P.}
\newblock \emph{\bibinfo{title}{Chebyshev and Fourier spectral methods}}
  (\bibinfo{publisher}{Courier Corporation}, \bibinfo{year}{2001}).

\end{thebibliography}

\end{document}